\documentclass{emulateapj}
\usepackage{apjfonts}
\usepackage{epsf}
\usepackage{amsmath}
\usepackage{graphicx}	
\bibpunct {(} {)} {;} {a} {} {,}	

\shortauthors{Ruhland et al.}
\shorttitle{The Structure of Sgr in BHB Stars}

\slugcomment{{\sc The Astrophysical Journal, in press}}

\begin{document}

%%%%% Added the \def\head{ lark.

%\def\head{

\title{The Structure of the Sagittarius Stellar Stream as traced by Blue
Horizontal Branch Stars}

\author{Christine Ruhland$^1$, Eric F.\ Bell$^2$,
Hans-Walter Rix$^1$, Xiang-Xiang Xue$^3$}

\affil{$^1$ Max-Planck-Institut f\"ur Astronomie, K\"onigstuhl 17, D-69117 Heidelberg, Germany; ruhland@mpia.de\\ 
$^2$ University of Michigan, Department of Astronomy, 830 Dennison Building, 500 Church St., Ann Arbor, MI 48105, USA \\
$^3$ Key Laboratory of Optical Astronomy, National Astronomical Observatories, CAS, 20A Datun Road, Chaoyang District, 100012, Beijing, China}

\begin{abstract}

We use a sample of blue horizontal branch (BHB) stars from the
Sloan Digital Sky Survey Data Release 7 to explore the structure
of the tidal tails from the Sagittarius Dwarf Galaxy.
We use a method yielding BHB star candidates with up to
$\sim$ 70\% purity from photometry alone.
The resulting sample has a distance precision of roughly 5\% and can probe distances in excess of 100 kpc.
Using this sample, we identify a possible extension to the trailing arm
at distances of 60$-$80 kpc from the Sun with an estimated significance of at least 3.8$\sigma$. Current models predict that a distant `returning' segment of the debris stream should exist, but place it substantially closer to the Sun where no debris is observed in our data. Exploiting the distance precision of our tracers,
we estimate the mean line-of-sight thickness
of the leading arm to be $\sim$3 kpc, and show that the two `bifurcated'
branches of the debris stream differ by only $1-2$ kpc in distance. With a spectroscopic very pure BHB star subsample, we estimate the
velocity dispersion in the leading arm, 37\,km\,s$^{-1}$, which is
in reasonable agreement with models of Sgr disruption.
We finally present a sample of
high-probability Sgr BHB stars in the leading arm of Sgr, selected to have
distances and velocities consistent with Sgr membership, to allow further
study.

\end{abstract}


%}%%%end head

%\twocolumn[\head]

\section{Introduction}

Tidal debris from dwarf galaxies and stellar clusters dissolving in the
Milky Way potential are an important contributor to the stellar halo of the
Milky Way \citep[e.g.,][]{SearleZinn1978,Ibata1994,Bullock2001,BullockJohnston2005,Belokurov2006,Bell2008}.  
In recent years, many elongated substructures
have been found in the stellar halo of the Milky Way
\citep[e.g.,][]{Ibata1995,Ibata2003,Yanny2003,Grillmair2006a,Grillmair2006b,Grillmair2006c,Belokurov2007}
and around other nearby galaxies such as Andromeda
\citep[e.g.,][]{Ibata2001a,McConnachie2009}, and a number of external
galaxies (e.g., NGC 891; \citealp{Mouhcine2010}; NGC 5907; \citealp{Zheng1999,
MartinezDelgado2008,MartinezDelgado2010}) showing that the build-up of stellar halos
through accretion of satellite galaxies is a common phenomenon.  Besides
the general implications such stellar satellite debris has for building and testing
the galaxy formation paradigm, the
detailed investigation of the individual structures provides important
information about the specific formation history of individual galaxies. The spatial distribution and kinematics of the
tidal debris of dwarf galaxies or globular clusters is also an important
source of information about the gravitational potential of the Milky Way
\citep[e.g.,][]{Johnston1999,Helmi2004a,Sag2MASS4,
Fellhauer2006,Koposov2009,LM2010a,Penarrubia2010a}.  

In this context, the Sagittarius stellar stream (Sgr), the most massive stellar stream around the Milky Way, is a central
case study. Discovered in 1994 \citep{Ibata1994}, the tidal
tail has been charted across more than one full wrap around the Milky Way
in
M-giants \citep[see also \citealp{Yanny2009}]{Sag2MASS1}, main sequence stars
\citep{Belokurov2006}, clusters \citep[e.g.][and references
therein]{Bellazzini2003}, and blue horizontal branch (BHB)
stars \citep{Newberg2003,Monaco2003,Clewley2006,Yanny2009,NiedersteOstholt2010}. The spatial tightness of the stream in combination with its full $360^{\circ}$ span makes it an important probe of the
potential \citep[e.g.,][]{Helmi&White1999,Moore1999,Ibata2001c,Ibata2002,Johnston2002,Sag2MASS3,Helmi2004a,Lewis2005,Binney2008}, of the disruption process
\citep{Ibata2001b,Helmi&White2001,Penarrubia2010b}, and of the impact of population
gradients and
cluster contents of the Sgr dwarf on the properties of the tail \citep[e.g.,][]{DaCosta1995,Sag2MASS1,MartinezDelgado2004,Bellazzini2003,LM2010b}.

Despite the wealth of observational data, models of the stream have failed
so far to match all the observational constraints by quite a margin. To explain the
observations different galaxy potentials have been invoked, with arguments for prolate
\citep{Helmi2004,Sag2MASS4}, spherical \citep{Fellhauer2006}, oblate
\citep{Sag2MASS3} or triaxial \citep{LM2010a} dark matter potentials. To explain some striking features, such as the `bifurcation' \citep{Belokurov2006},
\citet{Penarrubia2010b} invoked that the progenitor of the Sgr
stream may have been a rotating disk galaxy rather than a
pressure-supported dwarf galaxy as assumed by most previous models. However, no
single models seems to explain all parts of the stream while it is also not
entirely clear that all the overdensities found in the plane of the Sgr
stream are actually remnants of the same progenitor. A more precise and more complete empirical picture of the Sgr stream could be crucial in clarifying this issue, and this constitutes the central goal of the present paper.

In recent studies of the Sgr stream, there has been increased attention
toward BHB stars as a tracer population. Due to their
relative brightness they can be observed out to $\sim 100$ kpc in the
stellar halo of the Milky Way using Sloan Digital Sky Survey (SDSS) data. However, to take full advantage
of area coverage of surveys such as SDSS, the identification of these stars
needs to be done with photometric data alone. Many publications based their
selection on color boxes \citep{Yanny2000,Yanny2009,NiedersteOstholt2010}
that included a significant contamination from other blue stars (primarily
blue straggler (BS) stars).  Such contaminants can dominate in number, and are
1-2 mag fainter in absolute magnitude, confusing the interpretation of halo
structure using such samples.  

In this paper, we use SDSS data in the North Galactic Cap 
to study Sgr tidal debris. We choose
color-selected BHB star candidates as sparse tracers of the ancient, metal
poor
populations with well-defined absolute magnitudes, that are $\sim 3-4$
magnitudes brighter than the densely
populated main-sequence turn-off (MSTO) stars. Going beyond other recent
studies \citep[e.g.,][]{Yanny2009,NiedersteOstholt2010} of the Sgr system in
BHB stars we use a refined selection technique based on a spectroscopic
training sample which reduces the contamination by other stellar
populations (Bell et al. 2010). We show empirically that the
distance uncertainties in our sample are small, of the order of 5\%. We use
these stars to chart out the Sgr stream, focusing on three issues: delineating the distant ($>$ 50 kpc) overdensities that may be associated with
the Sgr trailing arm, on constraining and measuring
the thickness of the leading arm, and on presenting a sample of
high-probability Sgr
BHB star candidates with positions and velocities consistent with Sgr
membership for
further study. Furthermore we are explore the bifurcation that has been found by \citet{Belokurov2006} perpendicular to the orbital plane of the stream and its appearance in BHB stars.

\section{Data} \label{data}

\subsection{Blue Horizontal Branch Stars} 
\label{bhbselection}

\begin{figure}[t]
\begin{center}
\includegraphics[bb = 140 240 368 481, width=8.5cm]{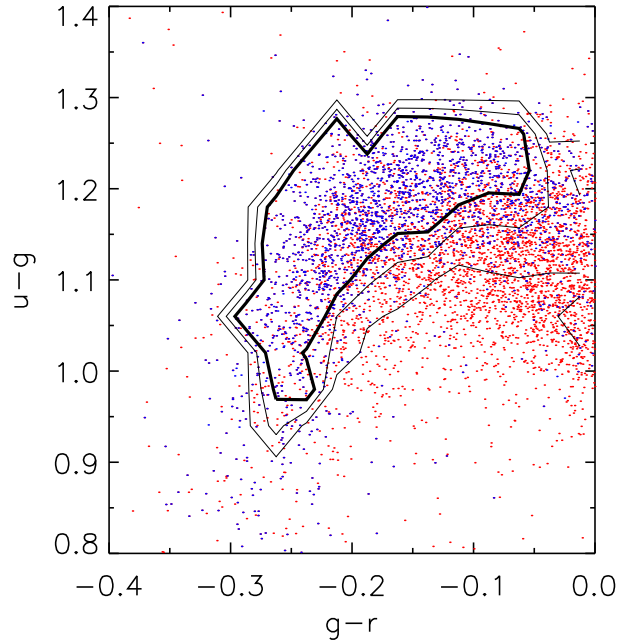}
\end{center}
\caption{\label{fig:selection}
Photometric BHB star selection. All candidate BHB stars shown have with spectroscopic classification \citep[following][]{Xue2008}. Stars spectrally classified as BHB stars are shown in blue; other stars are shown
in red. Contours show regions in $(u-g)$ vs. $(g-r)$ space where the fraction of targets classified as BHB stars exceeds 10\%, 30\% and 
50\%, respectively; the 50\% contour is thicker than the others. In what follows we focus primarily on BHB star candidates whose colors fall within the 50\% contour.}
\end{figure}

For this study, we use Data Release 7 \citep{SDSSDR7} of the SDSS to probe the Sagittarius stellar stream with BHB stars. 
The SDSS is an imaging and spectroscopic survey that has
so far mapped a little over $\sim 1/4$ of the sky. Imaging data are
produced simultaneously
in five photometric bands, namely, $u$, $g$, $r$, $i$, and
$z$~\citep{Fukugita1996,Gunn1998,Hogg2001,Gunn2006}. The data are processed through
pipelines to measure photometric and astrometric properties
\citep{Lupton1999,Stoughton2002,Smith2002,Pier2003,Ivezic2004,Tucker2006} and to select targets
for spectroscopic follow-up \citep{Blanton2003tiling,Strauss2002}.  

The horizontal branch is populated by stars which have developed past the main sequence stage and are now burning helium in their cores and hydrogen in the shell.
BHB stars have the dual advantages of a high luminosity (allowing probing
of the Milky Way halo to $>$ 100\,kpc), and have a small intrinsic spread in
absolute magnitudes.  Their main disadvantage is that the selection of a clean sample of BHB stars is challenging from photometry alone.  While broad cuts in
$u-g$ and $g-r$ are sufficient to isolate BHB stars and other A-type stars
(expected to be BS stars; \citealp{PrestonSneden2000},
\citealp{Sirko2004}) from low-redshift quasars and white dwarfs,
distinguishing BHB stars from the BS contaminants is
considerably more challenging \citep[e.g.,][]{Kinman1994,Wilhelm1999,Clewley2002,Sirko2004,Kinman2007,Xue2008,Smith2010}.  Previous
works have used broad color cuts designed to mitigate this contamination
(\citealp{Yanny2009} and \citealp{NiedersteOstholt2010} used the selection
in Figure 10 of \citealp{Yanny2000}; \citealp{Sirko2004} used a different cut
for their faint sample of BHB candidates).  Yet, these methods all suffer
from very substantial contamination from BS stars.  

Spectroscopy permits a fairly clean separation of BHB stars from BS stars on the
basis of surface gravity dependent Balmer line profiles.  \citet{Xue2008},
following 
\citet{Sirko2004}, use a two-stage cut to distinguish BHB from 
BS stars.  First, stars in the color box $0.8 < u-g < 1.6$ and
$-0.5 < g-r < 0.0$ with a relatively low line width and low flux in the
line core relative to the continuum are chosen (this reduces contamination
to $\sim50$\%).  Then, a S\'{e}rsic profile is fitted to the Balmer lines. 
By combination of these two criteria, a $>90$\% pure sample of BHB stars is
isolated.   Unfortunately, SDSS spectroscopy of BHB stars (mostly from SEGUE) is limited to certain areas of
sky, and only relatively bright BHB stars are targeted, meaning that BHB
stars more distant than $\gtrsim 50$\,kpc are not well-probed by the SDSS.  

Therefore, we have re-addressed the issue of photometric selection of BHB
star candidates (described in full in Bell et al. 2010).  We use the
spectroscopic classifications of $m_g <18$ stars from \citet{Xue2008} 
as a training set.  We
calculate the probability of a star in the color box $0.8 < u-g < 1.6$
and $-0.5 < g-r < 0.0$ being a BHB star from
this training set (Figure \ref{fig:selection}).
Blue data points 
show stars that are very likely to be BHB stars on the basis of their 
spectra (a contamination of much less than 10\% has been 
argued by \citealp{Xue2008} and \citealp{Sirko2004} for $m_g<18$).
The thick contour outlines the region of 
color-color space where the fraction of BHB candidates that 
are spectroscopically-classified BHB stars is $>50$\% 
and there were more than 16 stars in a bin of 
0.025$\times$0.04 mag.
Applying this selection to the SDSS DR7
\citep{SDSSDR7}, we obtain a candidate sample with 389,785 stars within the 
$0.8 < u-g < 1.6$, $-0.5 < g-r < 0.0$ color box. 
In the following we apply a lower
probability limit of 50\% for the photometric sample reducing the sample
size to 28,270 stars.  Tests show that this `$>50\%$' probability sample
isolates half of the $m_g < 18$ BHB star population, with a contamination
of
$20-30$\%.  Performance at fainter limits is expected to degrade gradually,
with increasing incompleteness and contamination (at $m_g \sim 20$ roughly
1/4 of BHB stars are expected to be kept, and contamination may be as
severe as 50\%; 
Bell et al. 2010). We will later test the influence of changing the
probability cuts (and therefore completeness/contamination) in Section \ref{probmap}.

\subsubsection{Kinematic Sample}

The radial velocity sample, which is a sub-sample of the photometric
sample, was selected based on the spectra as described above offering a
much higher BHB purity $\gtrsim 90$\% than the method applied on the stars with
photometry only. To not unnecessarily restrict the sample size, we use the
full radial velocity sample in these cases and ignore for these stars the
probabilities which were assigned based on their colors (i.e., we do not
use the lower probability limit of 50\% mentioned above). The total sample
size is 5233 stars, of which 807 are located in the Sgr plane (see Section \ref{planesection}). From these
807 stars 616 would fulfill the 50\% probability criterion, giving a
success rate of spectroscopic BHB stars in this selection of 76\%.
Throughout this paper the radial velocities are given in the Galactic standard of rest, which are the heliocentric radial velocities corrected for the Galactic rotation assuming a rotation velocity of 220\,km\,s$^{-1}$ for the local standard of rest and (+10.0,+5.2,+7.2)\,km\,s$^{-1}$ for the solar motion where the directions are defined as pointing towards the Galactic center, in the direction of rotation and towards the north Galactic Pole \citep[see][for details]{Xue2008}.

\subsubsection{Sagittarius in a Galactic Plane} \label{planesection}

For much of our analysis, we focus on stars in the presumed orbital plane of the Sgr stream only. We define this `Sagittarius plane' to encompass the
Sgr stream and the Galactic Center; this is presumably close to the orbital plane of the Sgr stream. 
To ensure consistency with models, we use the same pole as the Two Micron All Sky Survey (2MASS) papers \citep[e.g.][]{Sag2MASS1} at
$(l, b) = (273^{\circ}.8, -13^{\circ}.5)$.
Stars are considered to be in
the plane if they lie within $\pm \sim 27^{\circ}$ of this plane; this
definition naturally yields not a plane but a wedge, whose physical thickness increases with
distance from the Sun.  Stars are projected onto this plane by conserving
the distance to the Sun (i.e., the plane is a projection of shell segments
onto the plane).  
The Sagittarius plane defined here includes 73,066 stars (6905 with a BHB star
probability greater than 50\%) from the total 389,785 stars (28,270 with a BHB star probability
greater than 50\%) in the SDSS volume that are inside the color box.

\subsubsection{Empirical Distance Uncertainties}
\label{distuncert}

\begin{figure}[t]
 \centering
 \includegraphics[width=8cm,keepaspectratio=true]{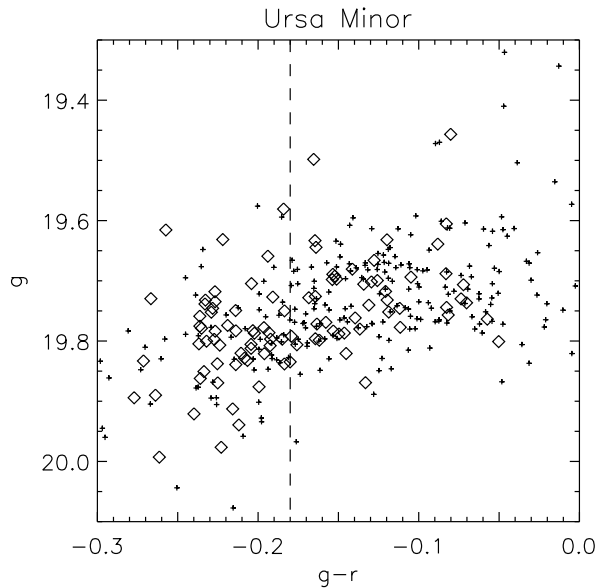}
 \caption{BHB section of the color magnitude diagram for the Ursa Minor dwarf spheroidal
galaxy. The diamonds represent the $ugr$-color selected BHB star candidates with $\geq$ 50\% BHB probability; the rest of the sample with lower probabilities is shown as crosses. It can be clearly seen that there is a trend
towards fainter magnitudes for bluer colors; the $g-r$ dependent $M_{g}$ calibration of \citet{Sirko2004} follows this trend closely. The vertical line shows the
position of the color cut applied to distinguish between red and blue BHB
stars.}
 \label{clustcmd}
\end{figure}

\begin{figure*}[th!]
 \centering
  \includegraphics[width=16cm,keepaspectratio=true]{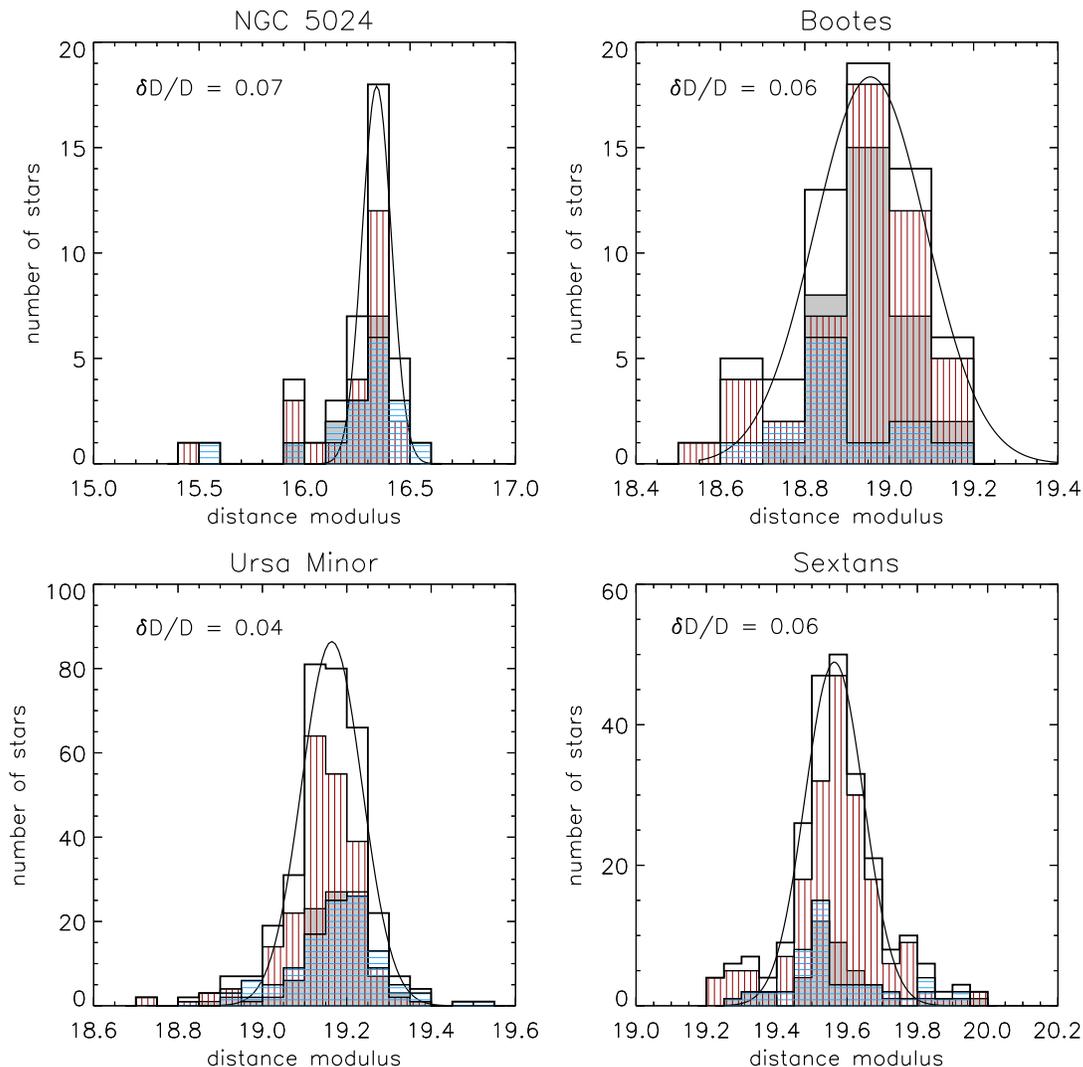}
 \caption{Distance accuracy for individual BHB stars. The panels show the distance modulus distribution of photometrically selected SDSS BHB stars in four clusters and dwarf
spheroidal galaxies. For each object, a Gaussian is fitted to the full sample
(unfilled histogram) to measure the width of the distribution as an
estimate of the distance uncertainty. The gray histogram shows only the
stars with a BHB probability greater than 50\% (Figure \ref{fig:selection}), the ones filled with vertical red and horizontal blue lines give the
distribution for the red and the blue part of the horizontal branch,
respectively. Clearly these clusters have a larger concentration of red BHB
stars. Results of the fit to both the $>50\%$ BHB probability sample and the red part are
shown in Table \ref{cluster_table}. The statistical distance uncertainty
$\delta D/D$ resulting from the fits to the full sample is also given in the plots,
being a bit higher for the full sample and a bit lower for the $>50\%$ sample than the expected value of $\delta D/D \sim 0.05$ \citep{Sirko2004}.}
 \label{clusthist}
\end{figure*}

\begin{table*}[th]
\begin{center}
\begin{footnotesize}
\caption{Distances to Clusters and Dwarf Spheroidals}
\begin{tabular}{cccccccc}
\hline \hline
Object & $\langle m-M \rangle_{prob>0.5}$ & $\sigma_{prob>0.5}$ & $\langle m-M \rangle_{red}$ & $\sigma_{red}$ & Literature
Values & $\delta D/D$ & Rel. Dist. Offset\\
& (mag) & (mag) & (mag) & (mag) & (mag) & & \\
\hline
NGC 5024 & 16.26 & 0.09 & 16.27 & 0.14 & 16.31\tablenotemark{a} & 0.04 & 0.02 \\ 
Bootes & 18.95 & 0.08 & 18.95 & 0.12 & 18.94 $\pm$ 0.14\tablenotemark{b} & 0.04 & 0.01 \\ 
Ursa Minor & 19.17 & 0.07 & 19.15 & 0.07 & 19.32 $\pm$ 0.12\tablenotemark{c} & 0.03 & 0.07 \\ 
Sextans & 19.56 & 0.12 & 19.57 & 0.11 & 19.75 $\pm$ 0.13\tablenotemark{d} & 0.05 & 0.06 \\ 
\hline\\
\end{tabular}
\centering
\tablecomments{Mean and standard deviation in distance modulus for four clusters
and dwarf galaxies given for a subsample with a BHB star probability $>50\%$ and for the red BHB stars
only. The second to last column gives the inferred distance uncertainty
$\delta D/D$ for the $>50\%$ sample of the color-selected BHB stars. The mean value is 0.04, which is a bit lower than the mean value of the full sample of 0.06. As literature values we give the mean values and standard deviations from a number of studies as listed. In
the last column, the relative distance offset is given which was calculated
using the literature values listed in the table.\\
\\
$^a$\citet{Harris1996paper}.\\
$^b$\citet{Belokurov2006b,DallOra2006,Siegel2006,DeJong2008}.\\
$^c$\citet{Mighell1999,Bellazzini2002,Carrera2002,Tammann2008}.\\
$^d$\citet{Matteo1995,Lee2003,Tammann2008}.}
\label{cluster_table}
\end{footnotesize}
\end{center}
\end{table*}

\begin{figure}[th]
 \centering
 \includegraphics*[bb=28 226 594 735, width=9cm]{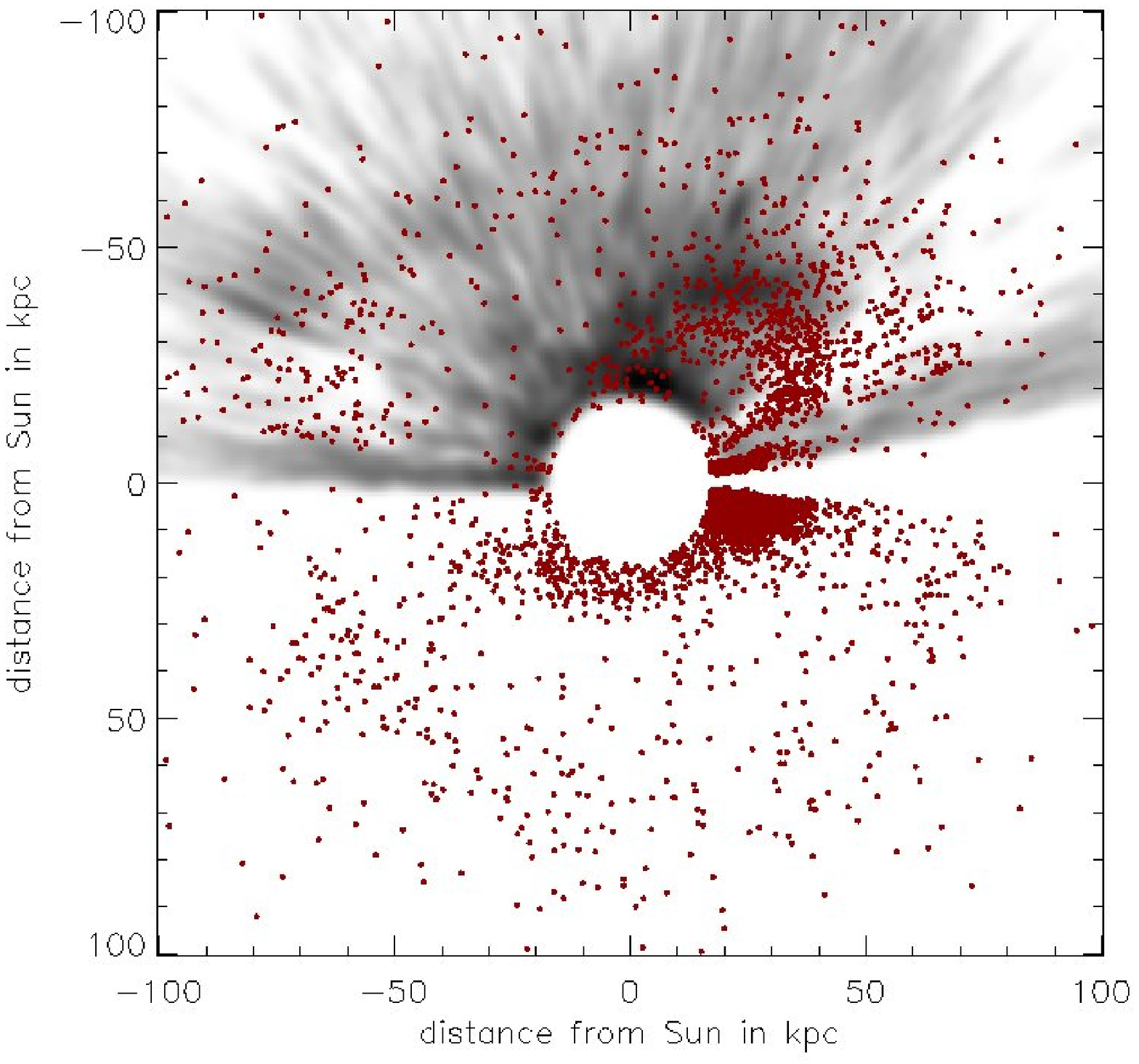}
 \caption{Distribution of M giants from 2MASS in the Sagittarius orbital plane is shown as red dots with a gray BHB star probability map in the background (see Section \ref{probmap}).
For both the BHB stars and the M giants we exclude the inner 20 kpc. The M giant population in the leading arm (+20,+35) appears closer to the Sun than the BHB stars, which we interpret as an 8\% mismatch in the distance scale between the two different populations.
Furthermore, the leading arm appears to be much more compact in width in BHB
stars than in M giants, presumably a reflection of the substantially larger distance
uncertainties in the M giants.}
 \label{planemgiants}
\end{figure}

As distance precision for the BHBs plays an important role for our analysis, we use several known globular clusters and dwarf spheroidals to determine
both the statistical and systematic uncertainties of the distance
determination\footnote{The distances were derived using the $g-r$-dependent $M_{g}$ calibration from Table 2 of \citet{Sirko2004} with [Fe/H]=$-1$ (different by less than the [Fe/H]=$-2$ calibration by $<$0.05 mag).}. 
A color-magnitude diagram (CMD) for one of the dwarf spheroidals is shown
in Figure \ref{clustcmd}. The typical shape of the blue horizontal branch
shows a nearly horizontal part at redder colors and a gradual trend towards
fainter magnitudes at the blue end, as can be seen in Figure
\ref{clustcmd}. Overall this trend causes an increase in the magnitude
spread and therefore distance measurement uncertainties toward bluer
colors. As we will show in Section \ref{probmap}, the Sagittarius
stream shows a larger concentration of `red' BHB stars. This indicates that
studying the red BHB stars separately can have two benefits compared to
looking only at the sample as a whole. {\it i)} The signal strength for the
stream will increase, and {\it ii)} the uncertainties introduced by the
deviations from the horizontal shape of the horizontal branch can be
reduced. 

Therefore, we divide the sample into a blue and a red part for further analysis. The $g-r$ value
at which we apply the cut throughout this paper is illustrated in Figure
\ref{clustcmd} by the vertical line. This cut is chosen to divide the
bright stars of the sample ($g < 18.5$ mag) in equally populated halves.
We determine the statistical error of the distance
measurement for BHB stars by measuring the spread of their distance moduli within one cluster (whose line-of-sight extent is negligible). The distance modulus distribution for the objects
is shown in Figure \ref{clusthist}. We fit Gaussians to these distributions
and use the standard deviation for estimating the statistical distance
uncertainty $\delta D/D$. We measure the mean value and the standard deviation for both the red part and
the $>50\%$ BHB probability sample (see Table \ref{cluster_table}). The distribution in
distance modulus of red and blue stars is also shown in Figure
\ref{clusthist} indicated by the blue and red shaded areas.
The results are shown in Table
\ref{cluster_table}. The mean statistical distance uncertainty for the
objects listed here is 4\% for the $>50\%$ sample and 6\% for the full sample.

Comparison with prior distance determinations (see Table \ref{cluster_table}) showed a systematic
underestimation of the distances in our results. This effect is of the order of 4\% in
distance, but also includes some variance which is probably also partly due
to the fact that the literature values were determined with different
methods.

With this test we cannot probe uncertainties in the distance determination that arise from a spread in metallicity. The metallicity-dependent BHB star models of \citet{Dotter2007} and \citet{Dotter2008} indicate a significant contribution to the distance uncertainties by a range of metallicities in the halo BHB stars. The overall uncertainty accounting for a combination of the scatter we see in single metallicity populations and the contribution of a scatter introduced by having a variety of metallicities is estimated to be less than 10\% in Bell et al. (2010).  In what follows we account only for the uncertainty which was estimated using single metallicity populations, which may underestimate the overall distance uncertainties (5\% vs. $<10\%$).

As a comparison data set we use M giants from the 2MASS \citep{Strutskie2006} to compare the distance scale of our BHB star data set in relation to other stellar populations, which were used for studying the Sgr stellar stream. In particular, this M giant data set was also used as the basis
for the models we will compare to later. M giants can be used as distance
indicators out to large distances making them a good stellar population for
studying the Sgr system (especially in the near infrared). Due to the
complete coverage of the sky, it is possible to observe the stellar stream
along its whole orbital path. A disadvantage of M giants as distance
indicators is their rather large distance uncertainties (argued to be
$\sim 17$\%; \citealp{Sag2MASS4}), and a likely distance offset with the BHB
and literature distance scales.

We derive a sample of M giants from the full 2MASS catalog following the method described in \citet{Sag2MASS1} for which we show the distribution in the plane of the Sgr stellar stream in Figure \ref{planemgiants}. The comparison with the BHB star population also shown in this plot reveals a distance offset between the two populations with the M giants being about $\sim 8 \%$ closer to the Sun than the BHB stars in the leading arm region of Sgr. As the distances of M giants are less well determined than those of BHB stars we also see a difference in the width of the leading arm in the different populations; the width seen in BHB stars is only $\sim 40\%$ of that seen in M giants as it appears in the samples presented here. Obviously this mismatch will propagate through to the models based on M giant observations \citep[e.g.,][]{Sag2MASS4,LM2010a}, so that we are expecting to see this mismatch to some degree in the comparison to these models. Note that we do not adjust our distance scale (or those of other data or models) to account for possible distance offsets in either case (the $\sim 4$\% mismatch between the BHB distance scale and the literature determinations, or the $\sim 8$\% mismatch between the BHB and M giant distance scales). In particular, this means that throughout the paper different data sets or models shown in the same plot can have different distance scales. Note that the offset to the distance scales of the two comparison samples have opposite directions, while the clusters and dwarf spheroidals have systematically larger distances, the M giants have smaller distances compared to our BHB star sample. This implies an even larger offset between these distance scales of about 12\%. Evidently a better characterization of the M giant distance scale would be of importance for a direct comparability of different stellar populations as distance indicators.

\subsection{N-Body Models for the Sgr Stream}

We will compare our BHB maps with simulations of the evolution of the
Sagittarius dwarf spheroidal in the Milky Way potential
\citep{Sag2MASS4,LM2010a,Penarrubia2010b} and summarize these models here. The \citet{Sag2MASS4} models adopt a smooth, rigid potential
representing the Milky Way, which consists of a Miyamoto-Nagai disk, a Hernquist
spheroid, and an axisymmetric logarithmic halo of different flattenings:  $q = 0.9$ (oblate), 1.0
(spheroidal), and 1.25 (prolate). We will also use a new model by
\citet{LM2010a} for comparison, which is based on a triaxial dark matter halo with a
minor/major axis ratio $(c/a)_\Phi = 0.72$ and a intermediate/major axis
ratio $(b/a)_\Phi = 0.99$ at radii $20 < r < 60$ kpc. This corresponds to a
nearly-oblate ellipsoid whose minor axis is contained within the Galactic
disk plane and approximately aligned with the line of sight to the Galactic
Center. In both model generations, the Sagittarius dwarf itself is represented by
$10^5$ self-gravitating particles. All of the models were constructed to
fit the system of the Sagittarius stellar stream as seen in 2MASS M giants.
To account for the photometric distance errors of the M giant sample, a
artificial random distance error of 17\% was applied to the simulated
debris particles. Following the suggestion of a triaxial halo, \citet{Penarrubia2010b} presented a model which does not assume a pressure-supported dwarf spheroidal galaxy as the progenitor of the Sgr stellar stream, but a late-type rotating disk galaxy. This model also reproduces a bifurcation in the leading arm of the stream as seen by \citet{Belokurov2006}.

\section{Results}

\subsection{Probabilistic BHB Density Maps} \label{probmap}

\begin{figure}[t]
 \begin{center}
 \includegraphics*[bb=121 236 353 745, width=7cm,keepaspectratio=true]{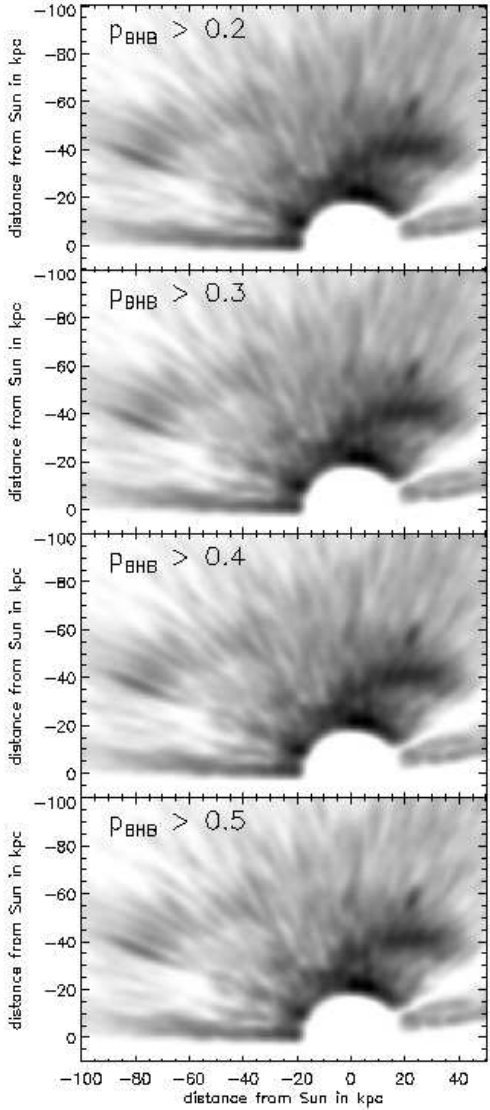}
\end{center}
 \caption{Density maps of BHB stars in a plane which includes the Sgr tidal
tail and the Galactic Center. The maps are derived as explained in Section
\ref{probmap} taking into account BHB star probabilities and distance
uncertainties. The four panels illustrate the dependence of these maps on the probability cut. The prominent overdensities in these maps are conserved for all probability cuts, showing that issues with the probability assignment are unlikely to affect our analysis significantly.}
 \label{prob}
\end{figure}

\begin{figure*}[t]
 \centering
 \includegraphics*[bb = 28 226 594 600, width=15cm,keepaspectratio=true]{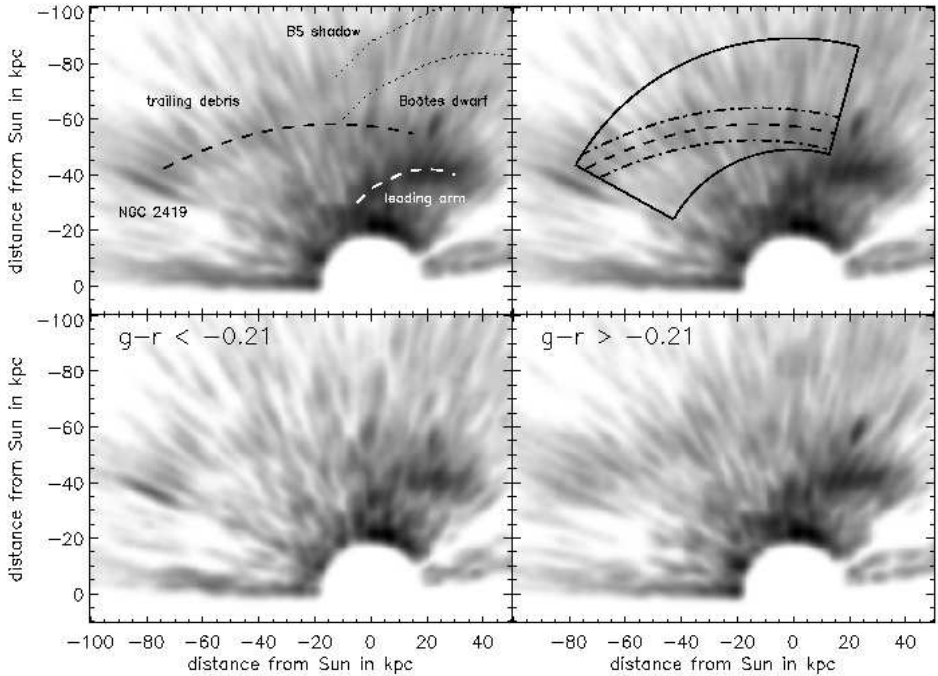}
 \caption{\footnotesize{Density maps of stars with $p_{BHB}>0.5$ in a plane which includes the Sgr tidal
tail and the Galactic Center. The maps are derived as explained in section
\ref{probmap} taking into account BHB star probabilities and distance
uncertainties, with a minimal accepted probability of 50\%. The upper panels
shows the full sample whereas the two lower panels show the sample split
into redder (right panel) and bluer (left panel) BHB stars. The Sun is
located at the origin. To enhance the contrast in
the regions of the Sgr debris the innermost 20 kpc are not shown. Clearly
visible is the leading arm of the Sagittarius stream stretching
approximately from $-$10 to 30 in x and $-$20 to $-$40 in y (indicated by
the short white dashed line). This feature is very clear in the full sample and
the red subsample whereas it is very patchy in the blue subsample. The two outer dotted black lines illustrate where the expected shadow of misinterpreted blue straggler stars should be assuming stars which are intrinsically 1.5 and 2 mag fainter, respectively. Another
clearly visible feature is the cluster NGC 2419 at (x,y)=($-$80,$-$35)
which is, in contrast to the leading arm, more prominent in the blue. It is not clear if NGC 2419 is associated with the candidate trailing arm (see Section \ref{comp_models}). We also see a very faint indication of an
overdensity in the region of $-$80 to 20 in x and around $-$50 in y,
spanning the region between the cluster and the leading arm (indicated
by the long black dashed line). If this overdensity is real it could possibly be
associated with Sgr and represent part of the trailing arm. The upper right panel illustrates the selection region used for an estimate of the significance of this overdensity discussed in Section \ref{probmap}. The Bo\"{o}tes dwarf galaxy can be clearly seen just above the leading arm of Sgr at (x,y)=(20,$-$60).}}
 \label{plane}
\end{figure*}

\begin{figure}[t]
 \begin{center}
   \includegraphics*[bb=141 256 566 651, width=9cm,keepaspectratio=true]{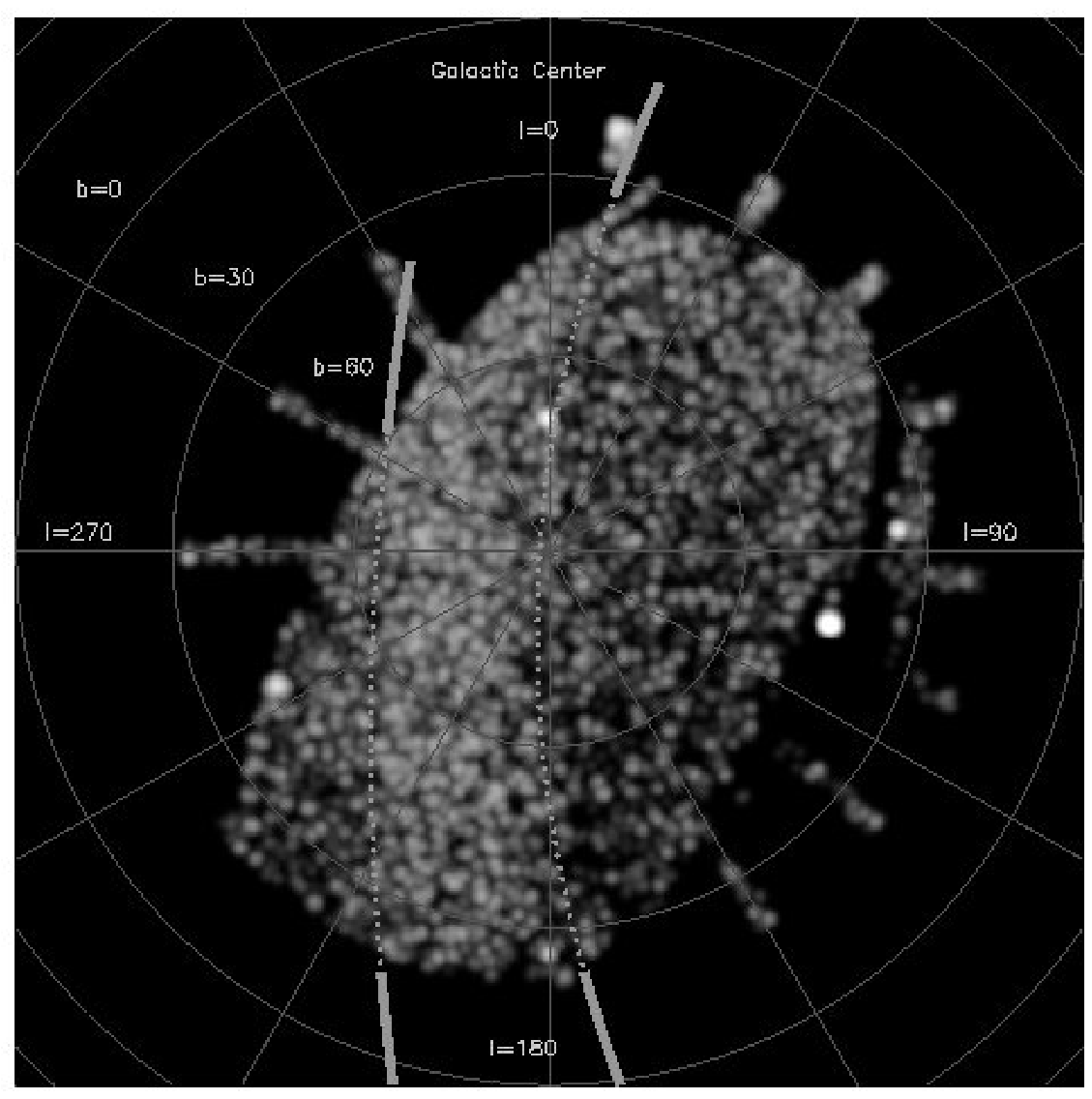}
 \end{center}
 \caption{On-sky view of a distance shell (60 kpc $< d <$ 80 kpc) which holds most part of the faint overdensity we see in the plane which could represent part of the trailing arm of Sgr. The borders of the plane are indicated by the dotted lines, for better visibility they are indicated by thick solid lines outside the survey field. The stars that we attribute to the Sgr trailing arm with $180^{\circ} < l < 270^{\circ}$ (and what we argue are blue stragglers in the leading arm at $270^{\circ} < l < 360^{\circ}$) are clearly confined to the Sgr plane.}
 \label{trailing}
\end{figure}

We then create maps to visualize the BHB star density, which account both for the finite probability that stars are BHB stars (as described in Section \ref{bhbselection}) and for the distance uncertainties.  We account for the distance uncertainty by viewing each star as an ensemble of 100 sub-objects, with line-of-sight distances drawn
from a Gaussian distribution with 5\% scatter (see Section
\ref{distuncert}) around the mean of the distance estimate for each individual
BHB star.  These sub-objects, each of which has a probability of 1\%, are
then multiplied by the probability they have to be a BHB star. Throughout this
paper, we only consider stars with $p_{BHB} > 0.5$ unless stated
otherwise. A map in the Sgr plane is then created by dividing the plane into
cells for which the probabilities of the included stars are summed.  The `signal' therefore depends on both spatial
abundance of stars and on probability of each to be a BHB star.  These maps
then get convolved with a Gaussian kernel with a size of $\sigma = 1$ kpc for presentation purposes\footnote{Later we will use a polar coordinate system, which is defined in Section \ref{thickness_section}, where a kernel of 0.5 kpc in distance and $\sim 2^{\circ}$ in the orbital angle coordinate is used.}.  
We apply this technique to create spatial maps of
the Sgr debris and for plotting the velocity distribution along the orbital
longitude of the system. In Figure \ref{prob}, we illustrate the effect of different probability cuts on the Sgr plane. This is of particular interest in the context reported in Section \ref{bhbselection} that the probability assignment is assumed to work less well for larger distances.

Our basic map, the distribution of BHB stars in the Sgr stream plane, is shown in Figure \ref{plane} (top panel). The uppermost panel
shows the full sample of stars, where the overdensities are pointed out by
dashed lines. Clearly visible is the leading arm of the stream to the right
of the plot (white line). Less prominent, but still significant (see below)\footnote{Among other features this appears somewhat more prominent if a lower probability cut for BHB stars is applied (e.g., 20\% or 30\%, see Figure \ref{prob}).}, is the overdensity denoted by the black dashed line. In common with \citet{Newberg2003} who detected part of the overdensity and \citet{Newberg2007} where it was also shown in BHB stars, we provisionally attribute this to the Sgr trailing arm. We find further support for this overdensity in the on-sky plot of a broad distance slice (60 kpc $< d <$ 80 kpc) covering most of the overdensity seen in the plane. Figure \ref{trailing} shows the on-sky view in which the plane is clearly visible as a overdense region.
Also clearly visible in Figure \ref{plane} is the globular cluster NGC 2419 at (x,y)=($-80$,$-35$), but its relation to the Sgr trailing stream is unclear.

In the direction of the leading arm, BS contamination is faintly visible as an echo of the leading arm at $\sim 80$\,kpc from
the Sun. It is noteworthy that our selection has significantly reduced this contamination compared to, e.g., \citet{NiedersteOstholt2010}.  Since we assign all stars with $p_{BHB}>0.5$ as BHB stars this causes an overestimate of BS star distances (by 1-2
mag, or a factor of two or so in distance, as observed). We illustrate the expected location of this shadow caused by stars in the leading arm region by the dotted black lines in Figure \ref{plane} (giving the transposition of the white line for stars overestimated by 1.5 and 2 mag, respectively). We show also the Bo\"{o}tes dwarf, which happens to lie in the Sgr plane.

We have adopted two different methods to estimate the significance of the candidate trailing stream. In the first approach, we estimate the significance in small areas of 4 kpc $\times$ $4^{\circ}$ along the trailing stream. We divide the plane into areas of constant radial and angular extent, and count the number of stars in these fields. For a field $i$ the number of stars in the field is $N_{i}$. The mean number of stars in a ring with constant heliocentric distance $\bar{N}_{ring}$ and standard deviation of $\sigma_{ring}$ is derived for each value of heliocentric distance range to account for the increasing volume of the wedge with increasing distance (see Section \ref{planesection} for a description of the geometry of the plane). We also exclude the angular range to the right of the area indicated in Figure \ref{plane} from the calculation of the mean and standard deviation for all distances to avoid the obvious overdensities from the leading arm in this area as well as the contamination at larger distances from misinterpreted BS stars. The significance of any deviation in the number of stars of each field within this sample of equidistant fields in units of the standard deviation $\sigma$ for region $i$ is given by $\sigma_{i}=(N_{i}-\bar{N}_{ring})/\sigma_{ring}$.

We take the region around the suggested position of the trailing arm as indicated in the upper right panel of Figure \ref{plane} by the dash-dotted lines and compare the average deviation of these fields with a comparison sample in the same plane but outside the trailing arm area. Note that this area does not include NGC 2419 to get a clean estimate of the significance of the proposed trailing arm. These fields are chosen in a way that the number of fields per distance interval of on- and off-stream fields is the same. The 57 on-stream fields show an average deviation of $+0.4 \sigma$ per field, indicating a weak overdensity, whereas the 57 off-stream fields show with an average deviation of $-0.6 \sigma$ per field the corresponding underdensity.

To get an idea of the significance of the whole extent of the structure we adopt a larger area, as shown in Figure \ref{plane} (upper right panel). Within this region consisting of 200 fields we randomly select a number of fields, equal to the number of stream fields we used earlier, and determine the average number of stars in this selection. Applied many times this bootstrapping method gives an estimate of the mean value and standard deviation we can expect in a randomly selected structure of this size. We find a mean value of 23.5 stars per field in the large box with a standard deviation of 1.3 stars. The average number of stars in the selected structure fields is 28.5 per field which corresponds to an deviation of 3.8 $\sigma$ from the mean value. The candidate stream fields are compared with all fields -- including stream fields -- potentially underestimating the significance.

In the two lower panels of Figure \ref{plane}, we show the maps that result after splitting the BHB sample in $g-r$ color at $g-r = -0.18$,
such that the number of stars with $g < 18.5$ is about equal in the red and blue subsamples.
The main motivation to do so is to probe the variations of the stellar
population in the Sgr stream. In Figure \ref{plane}, we show the red subsample
in the lower right panel and the blue subsample in the lower left panel.  
We find that Sagittarius (especially the leading arm) is much more
prominent in the red stars (see Figure \ref{plane}) while other parts, such
as NGC 2419, are dominated by blue stars.

In summary, we find Sgr's leading arm to be a prominent feature in BHB stars, even more so when the BHB star sample gets restricted to stars which are on the red part of the blue horizontal branch in $g-r$ color. Furthermore we observe a faint overdensity stretching out over most of the plane covered by SDSS, connecting the leading arm with the globular cluster NGC 2419. This overdensity was also described by \citet{Newberg2003,Newberg2007} as a part of the trailing arm of Sgr.

\subsection{Thickness of the Leading Arm, and Spatial Selection of Sgr BHB Star Candidates} \label{thickness_section}

\begin{figure}[t]
 \centering
 \includegraphics*[bb = 141 226 566 641, width=8.5cm,keepaspectratio=true]{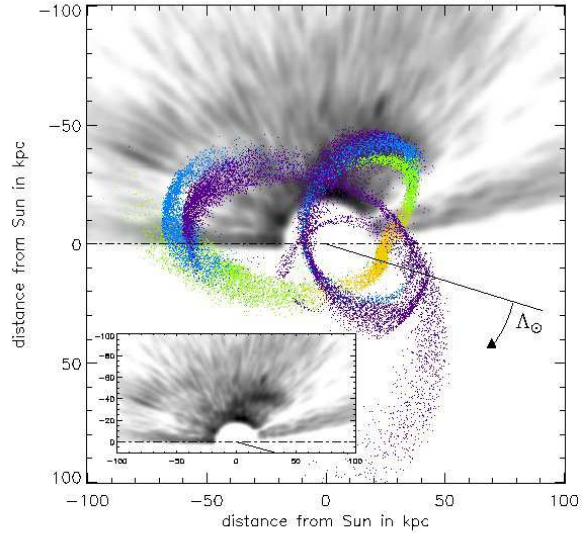}
 \caption{Comparison of the BHB star density ($>50\%$ sample) with the \citet{Sag2MASS4} prolate potential model of the
Sagittarius stellar stream and in
the small panel data alone for clarity. The colors showing the model
particles show debris lost from the progenitor during different epochs. The yellow points show debris stripped since the last apogalacticon, while green, blue, and purple show debris which became unbound two, three, and four orbits ago, respectively. The Sun is located in the center of the
coordinate system. The dash-dotted line gives the position of the Galactic
Plane, with the orientation is chosen such that it falls on the x-axis. The solid
black line indicates the direction to the Sgr dwarf galaxy which also
defined the zero direction of the longitudinal coordinate system, with the
angle $\Lambda_{\odot}$ increasing clockwise. To enhance the contrast in the
regions of the Sgr debris, the innermost 20 kpc are not shown. Except for the `leading arm', any data-model correspondence is not obvious.}
 \label{planemodel}
\end{figure}

\begin{figure*}[th]
 \centering
 \includegraphics*[bb = 28 210 594 490, width=12cm,keepaspectratio=true]{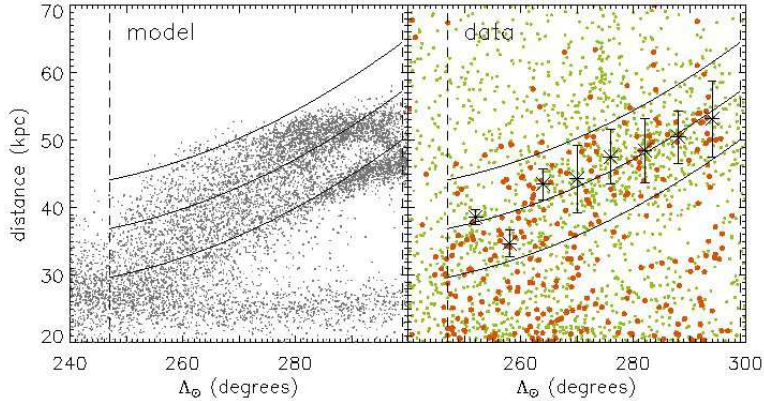}
 \caption{Heliocentric distance as a function of orbital longitude (see also the left column of Figure \ref{multipanel}). In the left panel, the prolate
model is shown as gray points. In the right panel, the data are shown with the small green
points representing the photometric sample (limited to BHB probabilities $>50\%$) and the larger orange points representing the
BHB sub-sample with radial velocities. The result of the Gaussian fit to the red BHB stars in the sample (Figure \ref{histfit}) is shown as asterisks (mean distance) and `error bars' (width $\sigma$). We fit the mean values with a second order polynomial
function (central line) and take $\pm 2 \sigma$ as our distance cuts (outer two lines). The dashed vertical lines mark the range in
$\Lambda_{\odot}$ over which the fit to the angle slices was performed. The
orange dots falling within these outlines we denote as the kinematic BHB selection for Figure \ref{veloselect1}.}
 \label{distselect}
\end{figure*}

\begin{figure}[h!]
 \centering
 \includegraphics[width=8cm,keepaspectratio=true]{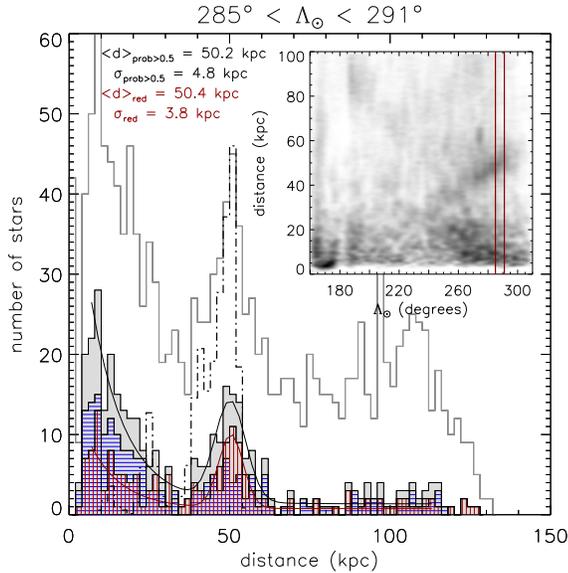}
 \caption{Distribution of heliocentric distances for $6^{\circ}$ wide
angle slice, $285^{\circ} < \Lambda_{\odot} < 291^{\circ}$, of Sgr's leading arm. The small panel on the upper right illustrates the angle slice
used for this histogram. The full sample is denoted by the empty solid line
histogram, while the gray histogram represents the sample with a $>$50\% probability
cut. A Gaussian has been fit to represent the stream stars in this subsample, while the fore-/back-ground has been fit by a power-law; the resulting distance and line-of-sight thickness are noted in the top left. Shown in the histogram filled with
red vertical lines is the subsample of red BHB stars and shown in the one filled with blue horizontal lines is the
subsample of blue BHB stars. The leading arm shows a slightly larger number of stars from the red subsample for all angles. The dash-dotted line shows
the histogram for the prolate \citet{Sag2MASS4} model in the same angle slice. It has been scaled
down to match approximately the number of objects in the data.}
 \label{histfit}
\end{figure}

\begin{table*}[th]
\begin{center}
\caption{Width of the Leading Arm}
\begin{tabular}{ccccccc}
\hline \hline
$\langle\Lambda_{\odot}\rangle$ & $\langle d \rangle_{prob>0.5}$ & $\sigma_{prob>0.5}$ &
$\sigma_{intr,prob>0.5}$ & $\langle d \rangle_{red}$ & $\sigma_{red}$ & $\sigma_{intr,red}$\\
(degree) & (kpc) & (kpc) & (kpc) & (kpc) & (kpc) & (kpc)\\
\hline
252 & 36.5 & 2.8 & 2.1 & 38.6 & 1.2 & - \\ 
258 & 35.6 & 1.3 & - & 34.7 & 2.0 & - \\ 
264 & 42.9 & 4.4 & 3.9 & 43.5 & 2.3 & - \\ 
270 & 46.7 & 6.2 & 5.7 & 44.3 & 5.0 & 4.2 \\ 
276 & 46.5 & 3.6 & 2.8 & 47.6 & 4.1 & 3.0 \\ 
282 & 48.5 & 5.0 & 4.4 & 48.5 & 4.8 & 3.7 \\ 
288 & 50.2 & 4.8 & 4.1 & 50.4 & 3.8 & 1.7 \\ 
294 & 52.7 & 6.5 & 6.0 & 53.1 & 5.6 & 4.4 \\ 
\hline\\
\end{tabular} 
\centering
%\hspace*{-1cm}
\tablecomments{Results for the Gaussian fit to the leading arm. Mean $\langle d \rangle$ and standard deviation $\sigma$
values are given both for the 50\% probability sample and for the red
subsample. We present also an estimate for the intrinsic width
$\sigma_{intr} = sqrt(\sigma_{obs}^2-(0.05\langle d \rangle)^2-\sigma_{bootstrap}^2)$.}
\label{width_table}
\end{center}
\end{table*}

In this section we measure the line-of-sight thickness of the
Sgr leading arm and use this measurement to select a sample of highly
likely Sgr member stars. In this subsection, we will present a selection based on the spatial distribution only which will be used for the analysis in the following subsection. Later we will restrict the selection of a `clean sample' to the radial velocity subsample for which we apply a similar selection technique. In what follows, we restrict our attention to the Sgr leading arm; the trailing arm (and candidate trailing arm debris) is in the wrong hemisphere and/or too distant to have SDSS radial velocity information. 

We adopt the heliocentric polar coordinate system defined by \citet{Sag2MASS1}
which was also used by \citet{Sag2MASS4} and \citet{LM2010a}. In this system, the angle is
defined as $\Lambda_{\odot} = 0^{\circ}$ passing through the main body of
Sgr and increasing along the direction of the trailing tail of Sgr. The
definition of the coordinate system is illustrated in Figure
\ref{planemodel} where the prolate version of the models is shown together
with our BHB star data in the large panel. The inset panel shows the data
alone.

We first measure the width of the leading arm using the full sample of
stars in the Sgr plane (not the Gaussian-distributed sub-objects), as is
appropriate for measuring line-of-sight distance scatter; the distribution of the stars is shown in Figure \ref{distselect}.  We divide the
angle-distance-plane into angular slices along an orbital angle range of
$250^{\circ} \lesssim \Lambda_{\odot} \lesssim 300^{\circ}$ and fit the
distance distribution of the stellar density with a function consisting of three
components: an exponential function and constant component to fit the
background distribution of halo stars, and a Gaussian for the Sagittarius
stream. The fit can be described by the expression\footnote{Due to the proximity of Bo\"{o}tes, we added a second Gaussian to this expression to isolate the profile of Sagittarius in the relevant bins from Bo\"{o}tes.} 
$\Sigma(d_{\odot}|\Lambda_{\odot}) = P_{0} + P_{1}*\text{exp}(\frac{-d_{\odot}}{P_{2}}) + P_{3}*\text{exp}(-0.5*(\frac{d_{\odot}-P_{4}}{P_{5}})^{2})$. The best fit
was determined using a chi-square algorithm. As the Sagittarius leading arm
is significantly more prominent in the red subsample of BHB stars we also
apply the fit to the red part alone (see Figure \ref{histfit}). For
comparison the histogram of the corresponding distribution in the models
(with a prolate potential) is shown by the dashed-dotted line. The
histogram is scaled down by a factor of seven to approximately match the
number of stars in the data. The models show a bifurcation of the leading
arm in distance between the debris lost in different orbits. We cannot see
this in our data, the relative separation and size of the peaks are roughly
of the same size as the fluctuations we see in the data in a typical angle
slice. The results of the Gaussian fit are shown in Table \ref{width_table}
and Figure \ref{distselect}; crosses denote the mean value and the `error' bars
show the standard deviation $\sigma$ around that mean.

We use these results as a first step in isolating a clean sample of BHB
stars.  A second-order polynomial is fit to the mean values, shown in Figure
\ref{distselect}.  We use this line, shifted by $\pm2$ times the mean
standard deviation as borders within which we select leading arm member
stars. In Section \ref{clean_section}, we will refine this selection by taking into account an additional selection in velocity space. In the following we will use the spatially selected sample defined here since the kinematic selection also very strongly limits the sample size to stars which have radial velocity data available. The spatially selected sample will be limited to BHB star probabilities greater than 50\%, whereas no probability cut is applied for the radial velocity sample since these stars are spectroscopically classified.

\subsection{Bifurcation of the Leading Arm Perpendicular to the Plane}

\begin{figure*}[th!]
 \centering
 \includegraphics*[bb = 14 216 580 792, width=12cm,keepaspectratio=true]{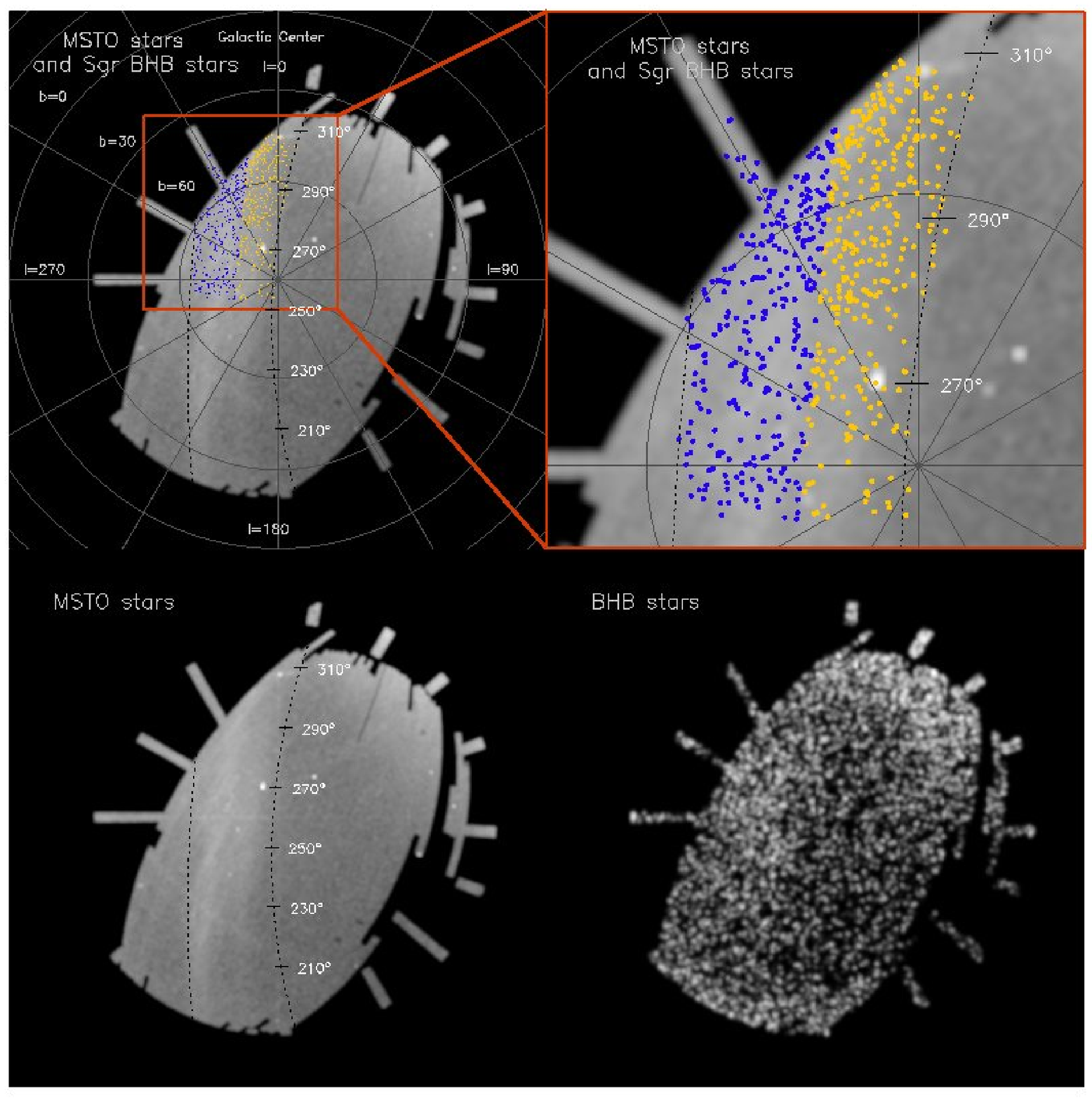}
 \caption{Map of the MSTO stars in a thick distance modulus slice, $16.5 \leq m-M \leq 17.5$, corresponding to a distance range of $20-32$ kpc (left side and top right corner). The dotted black lines indicate the borders of the plane as this projection represents a side-view of this plane. The orbital angle $\Lambda_{\odot}$ is given on the right of the plane segment. In the top panels, the BHB stars which were spatially selected as member stars of the leading arm are overplotted in yellow and blue. For better visibility, the area of interest (red box) is enlarged in the top right panel. The two colors denote the selection for the two parts of the arm following the appearance in MSTO stars. In the bottom right panel the BHB star distribution in the same distance shell is shown to illustrate why we can not select stars in the closer part of the stream where we see the strongest bifurcation in MSTO stars, as the BHB stars get much less abundant in the corresponding angle range (lower half of the plot).}
 \label{msto}
\end{figure*}

\begin{figure}[th!]
 \centering
 \includegraphics[width=8cm,keepaspectratio=true]{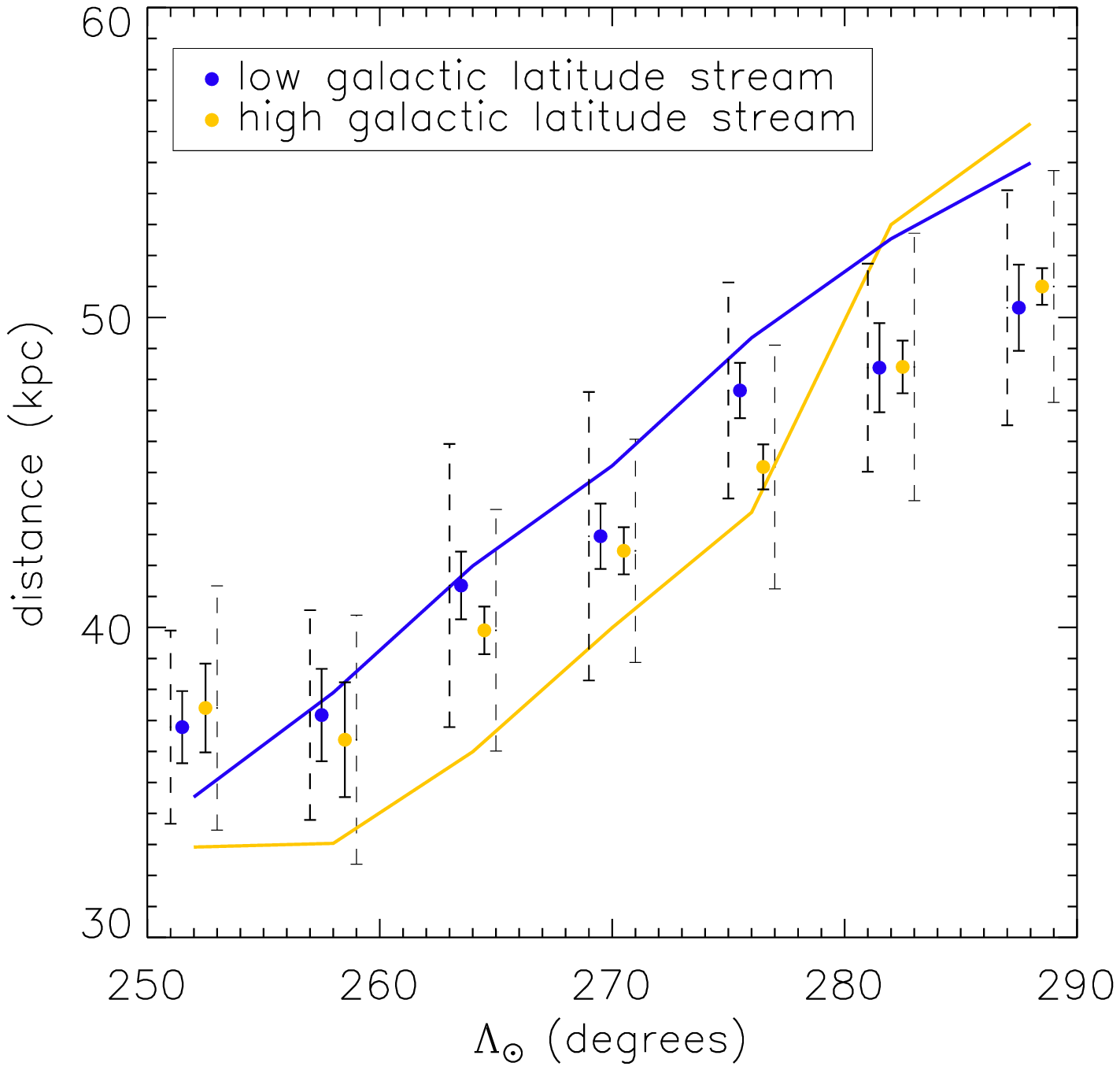}
 \caption{Mean distance and distance error corrected line-of-sight thickness of the two branches as determined from the spatially selected sample of Sgr BHB stars. Blue represents the branch at higher galactic latitude and yellow the one at lower galactic latitude (see also Figure \ref{msto}). For presentation purposes, the points are offset by half a degree to the left and right, respectively. The uncertainty of the mean value determination is determined by bootstrapping and given by the solid errorbars. The dashed `errorbars' represent the width of each substream, also corrected for the sample selection effects via bootstrapping. For better visibility, they are offset by another half a degree. The lines show the results of the same measurement on the \citet{Penarrubia2010b} models. We do not see a clear trend in the distance offset of the two branches as in the models and the observed offset is much smaller than predicted by the models and seen in other analyses with different stellar populations \citep[e.g.,][]{NiedersteOstholt2010}.}
 \label{bifurcation}
\end{figure}

Following \citet{Yanny2009}, we also look into the bifurcation of the leading arm as it was discovered by \citet{Belokurov2006}. When looking at thin distance slices of the SDSS using a population with a high abundance like MSTO stars one easily sees that the Sgr stream splits up into two parts. Given the relatively sparse distribution of BHB stars in the Sgr stream we use MSTO stars to define a selection for the two parts (see Figure \ref{msto}). In our BHB star sample itself we do not see any indication for a bifurcation. Following \citet{Bell2008}, we select MSTO stars in a color range of $0.2 < g-r < 0.4$ and a distance modulus range of $16.5 \leq m-M \leq 17.5$ assuming an absolute magnitude of $M_r=4.5$. This corresponds to a distance range of $20-32$ kpc. In Figure \ref{msto}, we show the distribution of these MSTO stars. In the upper panels, we also show the distribution of the Sgr BHB stars as spatially selected from Figure \ref{distselect}. Note that the clearly identifiable part of the leading arm in BHB stars is not in the region on the sky where the bifurcation is most apparent. In the lower right panel, we illustrate the low density of BHB stars in the relevant distance slice (same distance modulus selection as for MSTO stars) with $\Lambda_{\odot}\lesssim250^{\circ}$, which prevents us from investigating this part of the leading arm in BHB stars. Consequently, in the following we only study the leading arm for $\Lambda_{\odot}\gtrsim250^{\circ}$. In Figure \ref{bifurcation}, we present measurements of the mean and width of the two branches in thin angle slices. Note that in contrast to Figure \ref{distselect} this measurement was made on the pre-selected sample and not fitted to the data in the same fashion as illustrated in Figure \ref{histfit}. The two branches show similar distances with a 1-2 kpc variation in the mean distance values (see also Table \ref{bifurcation_table} for a listing of the results). Several studies showed a systematic separation in the distance of the two branches, such that the high galactic latitude part of the stream is closer for most of the leading arm as seen in the SDSS. \citet{Yanny2009} report this offset in the distance distribution of BHB stars along the leading arm by visual impression. The same trend was also seen by \citet{Belokurov2006} \citep[results listed in][]{NiedersteOstholt2010}, showing an offset of 2-3 kpc. An offset was also given by the \citet{Penarrubia2010b} models for which we show the mean distances of the two branches in Figure \ref{bifurcation}, separated and measured in the same way as our data. Although we do not see a clear separation in distances in our data, the mean distances of the two branches and their relation to each other are sensitive to small changes in the separation cut between the two branches.
Recently, \citet{Correnti2010} measured the distances of these two branches in Red Clump stars finding also only a small offset between them which is of a similar order as found in this study (see, e.g., their Figure 13).

\begin{table}[b]
\caption{Mean values with uncertainties and width for the two branches of the bifurcated stream as presented in Figure \ref{bifurcation}.}
\begin{tabular}{ccccc}
\hline \hline
$\langle\Lambda_{\odot}\rangle$ & $\langle d \rangle_{high}$ & $\sigma_{high}$ & $\langle d \rangle_{low}$ & $\sigma_{low}$\\
(degree) & (kpc) & (kpc) & (kpc) & (kpc)\\
\hline
252 & 37.4 $\pm$ 1.4 & 3.9 & 36.8 $\pm$ 1.2 & 3.1 \\ 
258 & 36.4 $\pm$ 1.9 & 4.0 & 37.2 $\pm$ 1.5 & 3.4 \\ 
264 & 39.9 $\pm$ 0.8 & 3.9 & 41.4 $\pm$ 1.1 & 4.6 \\ 
270 & 42.5 $\pm$ 0.8 & 3.6 & 42.9 $\pm$ 1.1 & 4.7 \\ 
276 & 45.2 $\pm$ 0.7 & 3.9 & 47.6 $\pm$ 0.9 & 3.5 \\ 
282 & 48.4 $\pm$ 0.9 & 4.3 & 48.4 $\pm$ 1.4 & 3.4 \\ 
288 & 51.0 $\pm$ 0.6 & 3.7 & 50.3 $\pm$ 1.4 & 3.8 \\ 
\hline
\end{tabular} 
\centering
\label{bifurcation_table}
\end{table}

\subsection{Kinematics and Comparison to Models} \label{comp_models}

\begin{figure*}[th!]
\centering
 \includegraphics*[bb = 20 170 550 750, width=16cm,keepaspectratio=true]{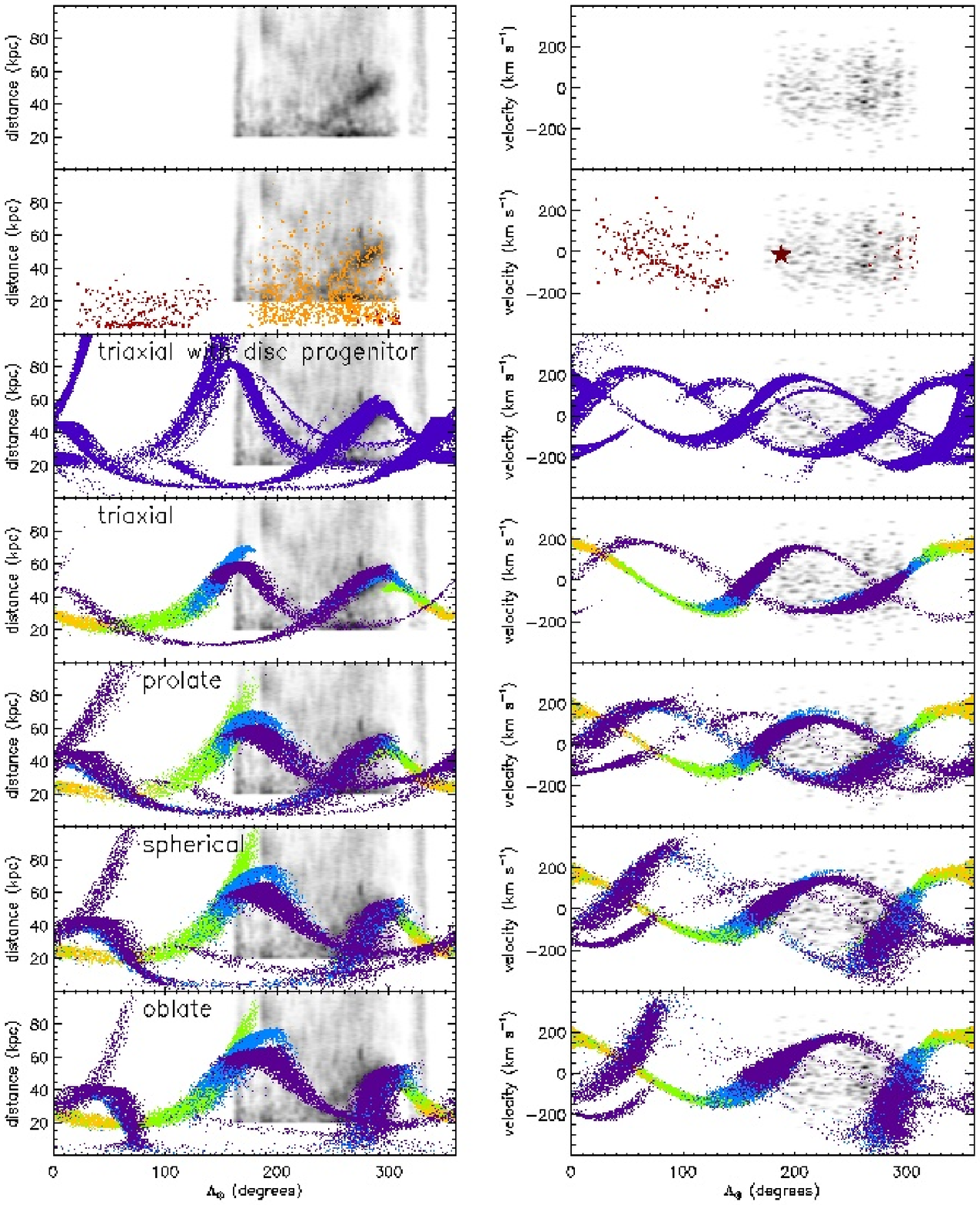}
 \caption{Distance and velocity of the Sgr stream for data and models as functions of orbital longitude. The upper row shows the data alone for stars within the Sgr orbital plane, analogous to Figure \ref{plane}. The right column shows the distribution of measured heliocentric
radial velocity in the Galactic standard of rest for the small subsample of
BHB stars with SDSS spectroscopy. These data are repeated in the background of each row. In the second row from the top we show a
sample of M giants with radial velocities from \citet{Sag2MASS2} in red (right, and the spatial distribution on the left), while we show the spatial distribution of our BHB star radial velocity sample in orange on the left. In the same row, the velocity of NGC 2419 is represented by the
dark red star. In the lower five rows, we show different models of the Sgr
debris. In the third and fourth rows, the triaxial halo models by \citet{LM2010a} and \citet{Penarrubia2010b} are shown. The fifth through seventh panels show the models by \citet{Sag2MASS4} using different halo potentials: prolate,
spherical, and oblate. The different colors in the models show debris from different orbits (see Figure \ref{planemodel}).
Comparing the BHB star maps with the models we see the best match for the leading arm region in the triaxial and prolate models for the spatial distribution. Although the radial velocity map does not resemble very well what is expected by the models, the main overdensities are also best covered by the triaxial and prolate models. Apart from the leading arm there is no good match between \textit{any} of the models and the data. The overdensities in BHB stars seen at distances between $\sim$ 50 and 90 kpc is not reproduced quantitatively by the models, nor do we see overdensities in the regions the models predict for the trailing arm.}
 \label{multipanel}
\end{figure*}

To improve our understanding of the origin of overdense regions of the Sgr
plane and the likelihood of those overdensities being associated with the Sgr
system, we compare our data with models of the Sgr debris by
\citet{Sag2MASS4}, \citet{LM2010a}, and \citet{Penarrubia2010b}. To complement our dataset with kinematic information, we
use a sample of radial velocities determined from SDSS DR7 using the method of \citet{Xue2008} and \citet{Xue2010}
(for an illustration of the spatial distribution of these stars with
kinematic information in the Sgr plane, see Figures \ref{multipanel} and \ref{distselect}). The models are based on the 2MASS data set of M giants, see Figure \ref{planemgiants} for their distribution. Figure \ref{multipanel} shows
probability maps made in the same fashion as described above for both
distance and velocity as a function of the orbital angle. In the three
lower panels, we overplot the models by \citet{Sag2MASS4}. From top to
bottom, they show the tidal debris in a prolate, spherical, and oblate
Galactic halo potential. In the third and fourth rows, two models using a triaxial halo potential are shown; in the fourth row the model from \citet{LM2010a}  and the model by \citet{Penarrubia2010b} in the third row which, in contrast to the other models, use a disk galaxy as a progenitor of the Sgr stellar stream.  In the second row, a radial velocity sample of M
giants \citep{Sag2MASS2} is shown alongside with our BHB star data set.
They cover mostly parts of the stream not covered by the SDSS.

In the following, we compare the location of the predicted debris in the different models with our observations of the distribution of BHB stars. This comparison is merely meant to illustrate tentative agreements and disagreements between the models and the data with the goal of identifying features and regions of interest for further investigation, and not give a conclusive answer for a best model.

The prolate and triaxial models, which are shown in the third to fifth rows of Figure \ref{multipanel}, clearly show the best consistency with the
leading arm which is the most prominent part in the SDSS BHB sample (at an orbital angle of $230^{\circ} \lesssim \Lambda_{\odot} \lesssim 300^{\circ}$ and heliocentric distance between 20 and 60 kpc).
On the other hand, the trailing arm from the \citet{Sag2MASS4} spherical and oblate models
stretches out to larger distances than the prolate model, qualitatively (but not quantitatively)
matching better the candidate Sgr debris shown in Figure
\ref{plane}\footnote{Part of the issue in reproducing such debris may be
related to the distance offsets between the BHB stars and M giant tracers
of the Sgr tail. The models were built to reproduce the smaller distances
characteristic of the M giant tracers; we speculate that models reproducing
better the leading arm in BHB stars would more easily yield a trailing arm
consistent with the distant candidate Sgr debris.}. The recent models by \citet{Penarrubia2010b} show a trailing arm which stretches out to much larger distances than in the other models. Still, we do not see a good match with the observed overdensity. 
\citet{Correnti2010} report detection of a trailing arm segment in Red Clump stars which appears to be consistent with the prolate models around the crossing region of the leading and trailing arm in the range of $220^{\circ} \lesssim \Lambda_{\odot} \lesssim 290^{\circ}$. This feature is observed at much smaller distances than what is suggested here. We do not focus on this distance range here, as in at least our investigation we find a high degree of contamination from and/or cross-talk with the Virgo overdensity. If the detection of \citet{Correnti2010} is interpreted correctly as part of the trailing arm the overdensity seen here could possibly belong to a different trailing wrap.

Turning to the possible association of NGC 2419 with the candidate trailing arm debris, we note that the heliocentric radial
velocity of NGC 2419 was measured by \citet{Peterson1986} to be $-20$\,km\,s$^{-1}$  which corresponds to a galactic standard of rest velocity of $-14$\,km\,s$^{-1}$ \citep{Newberg2003}. This corresponds well with the hypothesis
that the cluster is a part of the trailing stream near its apogalacticon
\citep{Newberg2003}. In addition, the properties of NGC 2419 are unusual
in its own right \citep[e.g.,][]{Dalessandro2008}: it is very luminous with
$M_V \sim -9.5$ and has a large half-light radius $r_h \sim 25$\,pc
\citep{Bellazzini2007}, placing it in a region of radius--luminosity
parameter space populated also by $\omega$\,Cen and M54, that have both
been argued to be the stripped cores of dwarf galaxies \citep[e.g.,][]{SarajediniLayden1995,Hilker2000,Romano2007,Bellazzini2008,Georgiev2009}.  
Yet, the situation with NGC 2419 in particular is not clear cut.  There is
no evidence of multiple stellar populations in NGC 2419 \citep{Cohen2010}, in apparent
contrast with the properties of, e.g., $\omega$\,Cen
\citep[e.g.,][]{Ripepi2007,Sandquist2008}.  Furthermore,
\citet{CasettiDinescu2009} have calculated a preliminary orbit for the
Virgo stellar overdensity, finding that it is very eccentric, and they
suggest that NGC 2419 may in fact be associated with the Virgo stellar
overdensity rather than Sgr.  Furthermore, we do not see a clear velocity
signature of trailing debris in the SDSS velocities (although it is unclear
if a signature is expected in the sparsely-sampled SDSS BHB velocity
data set).  Finally, the updated models of \citet{LM2010a} in a triaxial
potential show an {\it increased} inconsistency with NGC 2419 as described
by \citet{LM2010b}. 

Although the full velocity sample as we show it in this plot does not show
a very clear signal for the prominent leading arm, it is still obvious that
the main overdensities ($240^{\circ} \lesssim \Lambda_{\odot} \lesssim
300^{\circ}$) agree best with the models for the prolate and
triaxial versions. This will become clearer when we restrict the velocity
sample to stars within the region of the leading arm in distance space in
the next section.

A serious inconsistency with the models can be seen in the region where the
trailing arm is predicted to stretch into the region covered by the SDSS
(around ($-$60,0) and upwards in Figure \ref{planemodel}). In the data we do not see a signal which would
come anywhere near the intensity which is predicted by the models for this
part of the arm. The absence of such a counterpart indicates a serious
problem with the models. This can not be explained through differences in
the stellar populations in the debris; the models predict this part to
consist of stars that got unbound in the same orbits as the debris in the
part of the leading arm that we can observe in the SDSS.  We speculate that
this discrepancy may be alleviated in models tuned to reproduce better the
distances of the leading arm as traced by the BHB stars.

In the following section, we attempt to measure the
velocity spread of the Sagittarius stellar stream.  We continue our attempt to isolate a `clean' sample of stars most likely
belonging to the Sagittarius stellar stream. We use both positions and
kinematics to achieve a high reliability of our selection.  However, the
size and distribution of the radial velocity sample limit this selection
strongly.

\subsection{Selection of a `Clean Sample'
of BHB Star Candidates} \label{clean_section}

\begin{figure*}[t]
 \centering
\includegraphics*[bb = 28 210 594 509, width=12cm,keepaspectratio=true]{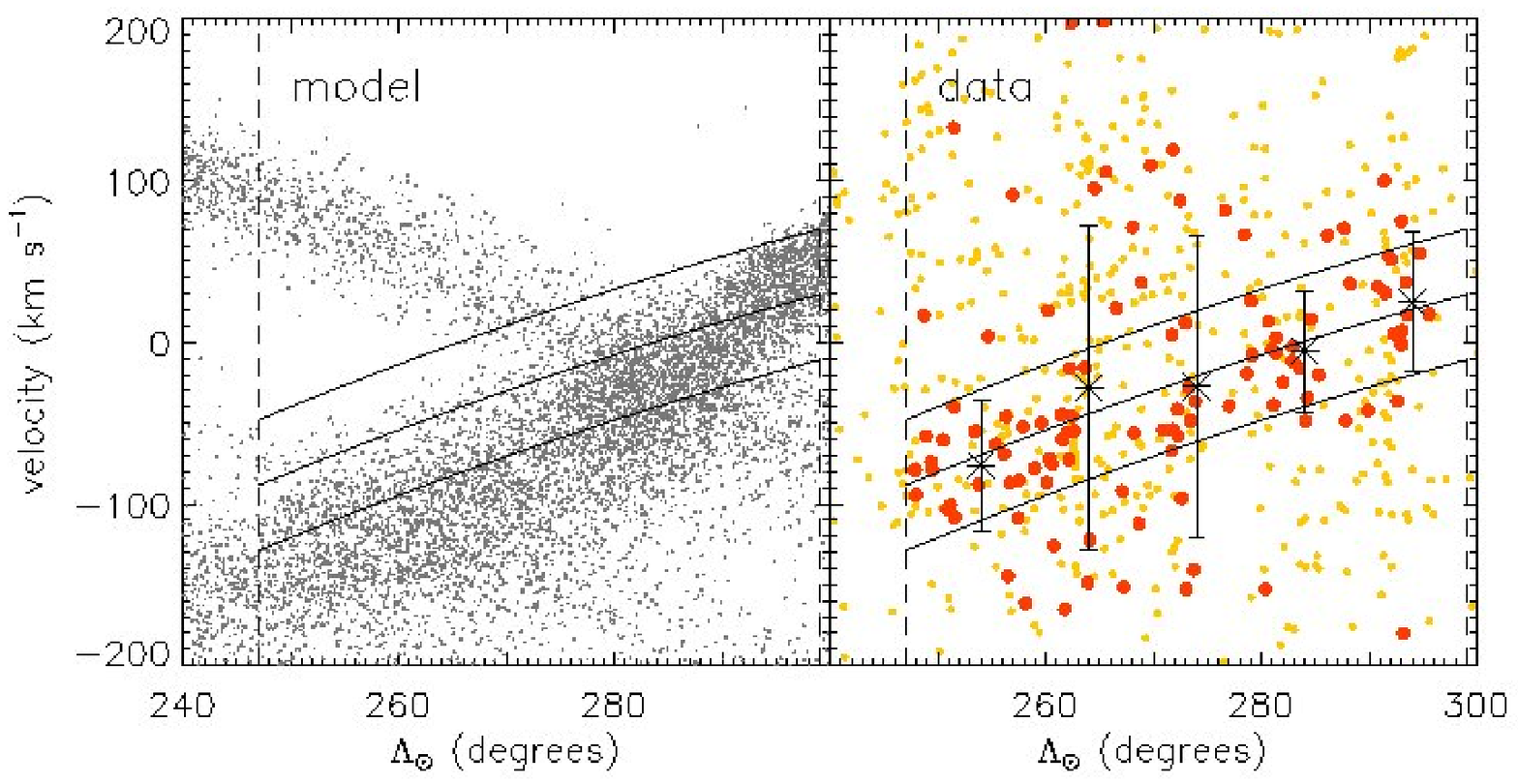}
 \caption{Heliocentric radial velocity in the Galactic standard of rest as a function of orbital longitude for the \citet{Sag2MASS4} prolate model and for BHB stars in the leading arm. The radial velocity
sample of BHB stars is shown by the orange and red dots (right panel); big red symbols denote
stars falling into the distance selection box of Figure \ref{distselect}
and small orange denotes everything falling outside this selection region. The
majority of red dots follow roughly the trend indicated by the models (left panel). We fit the velocity distribution of red symbols
with a Gaussian, using somewhat wider angle slices than before to account
for the sparser sample. The mean and standard deviation of these fits
are given by the asterisks and the bars (see also Table \ref{velo_table}). We apply a second order polynomial
fit to the mean values and shift this by the mean standard deviation in
both directions to identify a selection area for the `clean' sample.}
 \label{veloselect1}
\end{figure*}

\begin{table}[b]
\caption{Radial Velocities of the Leading Arm}
\begin{tabular}{ccccccc}
\hline \hline
$\langle\Lambda_{\odot}\rangle$ & $\langle$RVgal$\rangle$ & $\sigma_{obs}$ & $\sigma_{intr}$\\
(degree) & (km/s) & (km/s) & (km/s)\\
\hline
254 & -76.2 & 40.7 & 35.3 \\ 
264 & -28.4 & 100.7 & 98.3 \\ 
274 & -27.6 & 92.8 & 88.1 \\ 
284 & -5.7 & 37.3 & 35.8 \\ 
294 & 24.6 & 43.3 & 40.2 \\ 
\hline\\
\end{tabular} 
\centering
\tablecomments{Results for the Gaussian fit to the radial velocity distribution
of the leading arm sample. The intrinsic velocity spread is corrected for
the radial velocity error and the sample composition in the bin (via
bootstrapping).}
\label{velo_table}
\end{table}

In this section, we continue our effort to select a `clean sample' of Sgr BHB stars. In Section \ref{thickness_section}, we already made a spatial selection of the leading arm stars. In the following, we will restrict this selection to the radial velocity subsample to achieve a sample which follows the leading arm in both distance and velocity space. This selection of a
`clean sample' of Sgr BHB stars is based purely on the data, but agrees
qualitatively in both distance and velocity space with the Law et al.
models.

\begin{figure*}[th!]
\centering
 \includegraphics[bb = 28 210 594 509, width=12cm,keepaspectratio=true]{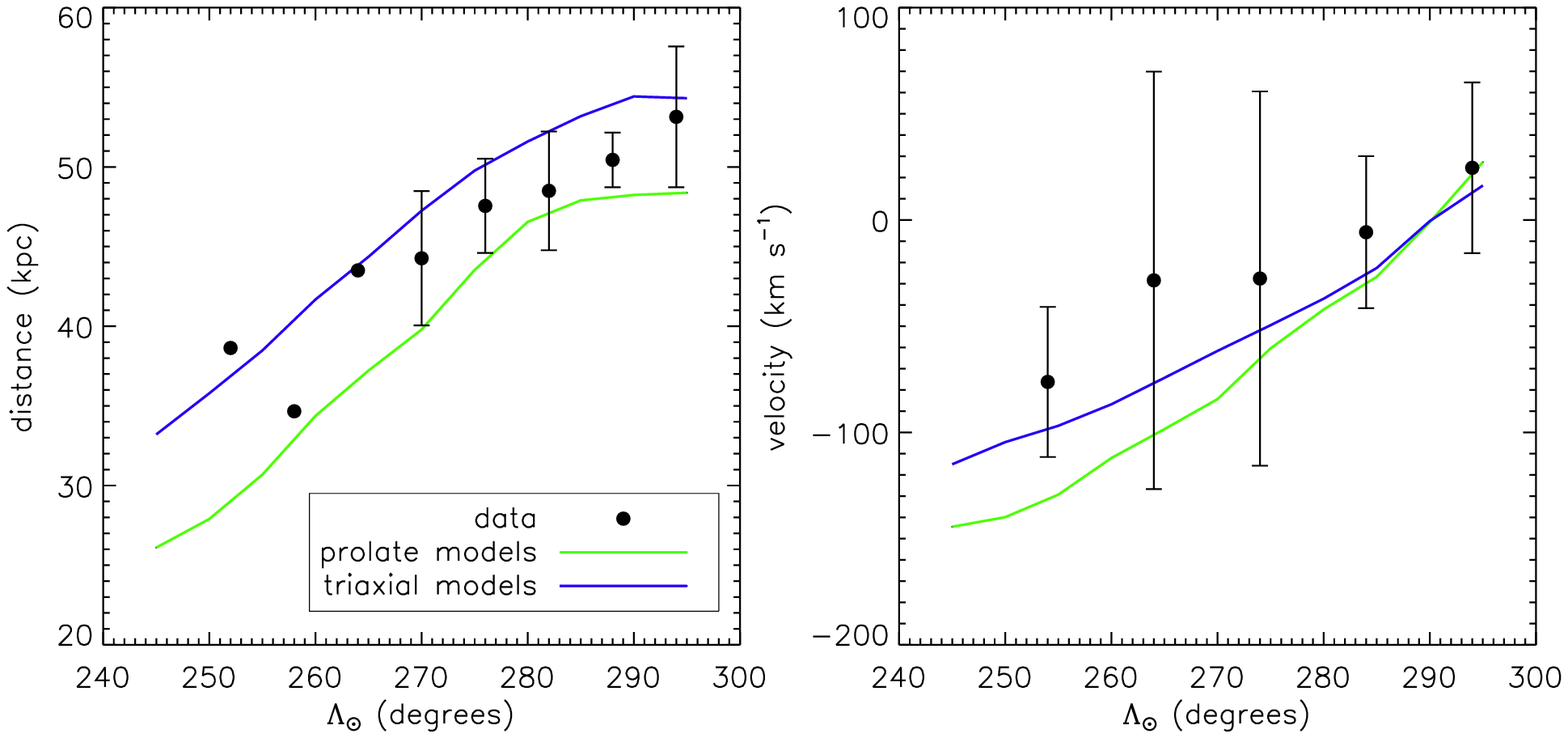}
 \caption{Distance and velocity dispersions of data from the spatial and kinematic selections for the leading arm and models. The velocities are again given in the Galactic standard of rest. We only show the models for a prolate and triaxial potential \citep{Sag2MASS2,LM2010a}
since these showed the best consistency with the properties of the leading
arm. The data points represent the mean and width measurements (bars) as presented in Figures \ref{distselect} and \ref{veloselect1}, showing a fit to the total sample on the left and the spatially selected sub-sample on the right. The data are corrected for distance uncertainties and for uncertainties
resulting from the sample selection in the bins by bootstrapping. For some bins these uncertainties are too large to resolve the intrinsic width of the stream.}
 \label{dispersion}
\end{figure*}

Figure \ref{veloselect1} shows the radial velocity full sample (orange), with
those lying in the distance selection in red.  As can be seen in comparison
with the models (in gray), the selected stars are mostly concentrated in an
area quite consistent with the general trend of the model. We isolate
candidate leading arm stars by taking angle slices in which we fit a
Gaussian to the distribution of the distance-selected stars (red).  Again
we fit the mean values with a second-order polynomial function (see Table
\ref{velo_table} for results). In Figure \ref{dispersion}, we show the
trends in distance and velocity, and their intrinsic dispersions, compared
with the models with prolate and triaxial potentials. Stars lying within
the distance selection, and within one mean standard deviation in either
direction of the velocity fit are included in our `clean' sample (see Table
\ref{cleansample_table} for a full listing of the objects). When using
this sample for further analysis one has to keep in mind that it is
strongly restricted by the uneven coverage and magnitude distribution of
the radial velocity sample.

\begin{table*}[th!]
\begin{scriptsize}
\caption{Sample selected for a high probability of association with
Sagittarius.}
\begin{tabular}{ccccccccccccc}
\hline \hline
R.A. & Decl. & $l$ & $b$ & $g$ & $u-g$ & $g-r$ & $p_{BHB}$ & HRV & HRVerr & RVgal &
Fe/H & Fe/H error\\
(deg) & (deg) & (deg) & (deg) & (mag) & (mag) & (mag) &  &
(km s$^-1$) & (km s$^-1$) & (km s$^-1$) & & \\
\hline
      149.31784 &       0.64158908 &        337.25807 &        59.124824 & 18.92 & 1.25 & -0.09 & 0.68 & -5.1 & 11.0 & -38.9 & -1.7 & 0.2\\
       149.32185 &       0.82575434 &        340.25544 &        57.966839 & 19.20 & 1.19 & -0.15 & 0.72 & 17.6 & 19.0 & -11.7 & -10.0 & -10.0\\
       151.66303 &       0.71303259 &        332.93978 &        60.252136 & 18.70 & 1.18 & -0.06 & 0.44 & 65.9 & 11.0 & 25.7 & -1.4 & 0.1\\
       151.68573 &       0.83859047 &        336.16948 &        59.330910 & 19.52 & 1.05 & -0.30 & 0.75 & 48.5 & 24.0 & 12.9 & -10.0 & -10.0\\
       152.46681 &       0.72481130 &        354.09098 &        51.794464 & 19.15 & 1.21 & -0.12 & 0.70 & 32.8 & 8.0 & 30.3 & -1.2 & 0.1\\
       152.73950 &       0.80837915 &        355.95736 &        50.489586 & 19.11 & 1.15 & -0.19 & 0.61 & 5.2 & 21.0 & 7.0 & -10.0 & -10.0\\
       162.91279 &       0.63505894 &        352.51890 &        50.663479 & 19.12 & 1.30 & -0.17 & 0.05 & 11.9 & 8.0 & 5.2 & -1.9 & 0.0\\
       163.11225 &       0.80012300 &        355.61411 &        48.732353 & 19.11 & 1.20 & -0.02 & 0.29 & 54.4 & 18.0 & 55.0 & -10.0 & -10.0\\
       165.03563 &       0.82406409 &        291.64879 &        62.180853 & 18.84 & 1.29 & -0.13 & 0.05 & 73.4 & 16.0 & -16.2 & -10.0 & -10.0\\
       165.26730 &       0.67366782 &        294.87797 &        62.247052 & 18.87 & 1.28 & -0.02 & 0.05 & 71.1 & 7.0 & -15.7 & -1.9 & 0.1\\
       171.35795 &       0.77705579 &        334.88970 &        58.104219 & 18.61 & 1.00 & -0.02 & 0.09 & 42.4 & 11.0 & 2.8 & -2.4 & 0.4\\
       177.49904 &       0.69659886 &        355.65421 &        49.925343 & 18.97 & 1.18 & -0.19 & 0.81 & 15.9 & 14.0 & 16.8 & -10.0 & -10.0\\
       181.41167 &       0.68832934 &        292.43919 &        63.422670 & 18.86 & 1.20 & -0.14 & 0.80 & 12.5 & 16.0 & -72.5 & -10.0 & -10.0\\
       181.54416 &       0.66210870 &        314.60980 &        63.437217 & 19.01 & 1.12 & -0.13 & 0.41 & 7.7 & 11.0 & -54.4 & -1.5 & 0.1\\
       181.57671 &       0.68079996 &        316.52701 &        63.065209 & 18.76 & 1.10 & -0.19 & 0.34 & 6.4 & 7.0 & -54.1 & -1.7 & 0.0\\
       185.50090 &       0.76998992 &        332.59194 &        60.542340 & 18.72 & 1.22 & -0.19 & 0.87 & 20.9 & 17.0 & -19.4 & -10.0 & -10.0\\
       185.50929 &       0.83180709 &        333.63472 &        60.321873 & 18.94 & 1.22 & -0.14 & 0.80 & 30.8 & 16.0 & -8.0 & -10.0 & -10.0\\
       185.13369 &       0.79081299 &        342.06521 &        57.760352 & 19.08 & 1.26 & -0.06 & 0.58 & 10.2 & 18.0 & -15.6 & -10.0 & -10.0\\
       173.32034 &        1.2211629 &        359.16254 &        50.629827 & 19.09 & 1.24 & -0.06 & 0.61 & 27.3 & 15.0 & 37.1 & -10.0 & -10.0\\
       191.92214 &        1.1856847 &        353.74592 &        52.452357 & 19.20 & 1.25 & -0.19 & 0.05 & 37.6 & 8.0 & 34.4 & -1.7 & 0.5\\
       192.82045 &        1.1740127 &        342.24188 &        58.467647 & 19.08 & 1.23 & -0.15 & 0.90 & 22.3 & 18.0 & -2.5 & -10.0 & -10.0\\
       196.78478 &        1.0952469 &        353.51720 &        55.082748 & 19.13 & 1.16 & -0.10 & 0.49 & 39.4 & 15.0 & 36.4 & -10.0 & -10.0\\
       249.77752 &      -0.12263493 &        343.67858 &        60.877749 & 18.87 & 1.26 & -0.05 & 0.05 & 15.5 & 14.0 & -4.3 & -1.3 & 0.0\\
       236.23186 &       0.32725518 &        349.53368 &        58.140268 & 19.17 & 1.04 & -0.14 & 0.04 & 24.6 & 18.0 & 14.3 & -10.0 & -10.0\\
       247.67761 &       0.34199587 &        297.76277 &        68.420754 & 18.43 & 1.25 & -0.14 & 0.89 & 19.0 & 13.0 & -45.9 & -2.2 & 0.1\\
       195.51961 &      -0.73406591 &        258.01091 &        71.848421 & 18.38 & 1.18 & -0.18 & 0.81 & 4.2 & 12.0 & -58.2 & -2.4 & 0.2\\
       198.27333 &      -0.68782645 &        254.48333 &        72.188679 & 18.36 & 1.05 & -0.21 & 0.30 & -34.2 & 11.0 & -94.5 & -1.3 & 0.6\\
       203.08792 &      -0.74637791 &        261.17467 &        74.794533 & 18.31 & 1.15 & -0.11 & 0.16 & -8.6 & 9.0 & -60.4 & -1.7 & 0.2\\
       203.46564 &      -0.74396165 &        294.52133 &        78.208821 & 18.70 & 1.31 & -0.13 & 0.05 & -18.4 & 11.0 & -52.4 & -1.7 & 0.3\\
       235.28992 &      -0.66049236 &        338.03357 &        68.100269 & 18.99 & 1.22 & -0.19 & 0.87 & -27.2 & 8.0 & -48.5 & -1.9 & 0.2\\
       237.33833 &      -0.64090826 &        336.47791 &        68.325007 & 19.21 & 1.29 & -0.13 & 0.05 & 35.0 & 8.0 & 11.9 & -1.4 & 0.2\\
       183.15324 &      -0.35575570 &        341.09162 &        68.226778 & 18.43 & 1.02 & -0.01 & 0.09 & -19.8 & 6.0 & -36.7 & -1.3 & 0.1\\
       184.33572 &      -0.30121489 &        359.67778 &        59.904530 & 18.98 & 1.09 & -0.21 & 0.63 & -44.8 & 7.0 & -34.2 & -1.9 & 0.1\\
       186.20173 &      -0.34449501 &        338.42710 &        69.530976 & 18.77 & 1.22 & -0.17 & 0.90 & -38.8 & 5.0 & -57.8 & -1.6 & 0.0\\
       230.36425 &      -0.33667675 &        312.91600 &        77.668905 & 18.65 & 1.16 & -0.08 & 0.41 & -33.5 & 12.0 & -60.2 & -1.8 & 0.2\\
       231.31942 &      -0.34046227 &        350.71450 &        70.209927 & 18.50 & 1.17 & -0.05 & 0.44 & -25.8 & 11.0 & -28.0 & -1.9 & 0.1\\
       247.71507 &      -0.35833233 &        303.66505 &        76.708831 & 18.62 & 1.15 & -0.20 & 0.66 & -37.3 & 13.0 & -72.1 & -1.6 & 1.1\\
       177.38054 &       0.15937633 &        281.06974 &        77.423385 & 18.41 & 1.21 & -0.11 & 0.70 & -22.3 & 10.0 & -63.0 & -2.4 & 0.4\\
       213.82899 &      0.048201690 &        284.98406 &        76.767400 & 18.51 & 1.26 & -0.21 & 0.82 & -3.8 & 12.0 & -46.0 & -1.8 & 0.5\\
       213.89940 &       0.18624951 &        300.73988 &        77.134295 & 18.58 & 1.16 & -0.01 & 0.29 & -15.0 & 9.0 & -49.9 & -1.3 & 0.3\\
       214.76100 &       0.13294046 &        319.00435 &        77.535698 & 18.52 & 1.28 & -0.25 & 0.05 & -31.7 & 12.0 & -54.9 & -1.3 & 0.5\\
       12.739674 &        15.849297 &        324.89192 &        85.302086 & 18.74 & 1.20 & -0.22 & 0.79 & -82.7 & 15.0 & -85.5 & -10.0 & -10.0\\
       21.239977 &        14.300256 &        293.68604 &        81.423698 & 18.51 & 1.27 & -0.14 & 0.89 & -64.0 & 9.0 & -87.0 & -1.0 & 0.1\\
       11.571439 &        15.267388 &        322.21782 &        83.117026 & 18.43 & 1.15 & -0.18 & 0.61 & -69.6 & 5.0 & -78.0 & -2.3 & 0.1\\
       37.994698 &       -9.4446953 &        322.70253 &        79.105115 & 18.51 & 1.19 & -0.18 & 0.81 & -38.3 & 9.0 & -55.5 & -1.8 & 0.0\\
       62.479321 &       -6.3200403 &        256.52080 &        75.094461 & 18.11 & 1.07 & -0.13 & 0.04 & -24.1 & 8.0 & -74.1 & -2.0 & 0.0\\
       40.554837 &       -8.7054956 &        260.66485 &        77.113803 & 17.92 & 1.15 & -0.15 & 0.41 & -60.3 & 7.0 & -103.2 & -1.8 & 0.1\\
       42.311590 &       -8.4109077 &        260.92325 &        78.271266 & 18.15 & 1.19 & -0.16 & 0.72 & -62.2 & 6.0 & -100.7 & -1.8 & 0.1\\
       181.39053 &       -3.3742984 &        279.90974 &        67.307502 & 18.24 & 1.21 & -0.21 & 0.67 & 9.1 & 9.0 & -69.2 & -1.7 & 0.1\\
       199.43635 &       -2.8485374 &        298.10400 &        74.487378 & 18.30 & 1.21 & -0.13 & 0.80 & -42.1 & 9.0 & -87.0 & -2.0 & 0.1\\
       170.96747 &       -2.5415740 &        302.82768 &        75.868349 & 18.31 & 1.26 & -0.16 & 0.91 & -37.5 & 8.0 & -75.4 & -1.5 & 0.1\\
       192.16892 &       -2.2424141 &        273.75164 &        68.019640 & 18.32 & 1.12 & -0.24 & 0.70 & -10.9 & 11.0 & -88.1 & -1.0 & 0.2\\
       174.28068 &       -3.2210852 &        335.45465 &        68.964352 & 19.03 & 1.23 & -0.23 & 0.83 & -17.8 & 7.0 & -41.4 & -1.7 & 0.1\\
       181.68676 &       -3.2304553 &        301.78123 &        73.471968 & 18.58 & 1.21 & -0.07 & 0.61 & 1.2 & 3.0 & -44.9 & -1.7 & 0.1\\
       184.50026 &       -2.7444692 &        261.08378 &        79.380844 & 18.25 & 1.10 & -0.02 & 0.07 & -5.6 & 11.0 & -39.8 & -1.7 & 0.1\\
       185.92025 &       -2.8228684 &        245.46876 &        77.417677 & 17.96 & 1.22 & -0.15 & 0.90 & -40.7 & 3.0 & -79.2 & -1.8 & 0.1\\
       188.14379 &       -2.8617039 &        245.13008 &        77.769645 & 18.28 & 1.21 & -0.15 & 0.80 & -42.0 & 4.0 & -79.1 & -1.4 & 0.1\\
       187.30791 &       -2.7770480 &        249.11382 &        79.733714 & 18.85 & 1.18 & -0.04 & 0.30 & -48.0 & 20.0 & -79.0 & -10.0 & -10.0\\
       188.69635 &       -2.7577484 &        259.84128 &        80.889018 & 18.16 & 1.15 & -0.17 & 0.42 & -80.2 & 8.0 & -108.5 & -1.9 & 0.2\\
       179.19443 &       -2.3958845 &        351.97783 &        50.812086 & 19.50 & 1.17 & -0.27 & 0.68 & 59.3 & 13.0 & 51.3 & -1.0 & 0.3\\
       179.27744 &       -2.4873776 &        353.19843 &        50.061905 & 19.38 & 1.17 & -0.20 & 0.81 & 3.7 & 11.0 & -1.5 & -2.0 & 0.2\\
       182.51252 &       -2.2918592 &        326.21682 &        60.272876 & 18.84 & 1.22 & -0.17 & 0.90 & 12.4 & 12.0 & -39.3 & -1.8 & 0.3\\
       187.04228 &       -2.3998936 &        335.66563 &        58.430511 & 19.04 & 1.14 & -0.20 & 0.66 & 31.2 & 13.0 & -6.5 & -10.0 & -10.0\\
       188.39442 &       -2.3430983 &        306.23529 &        62.775039 & 18.81 & 1.18 & -0.08 & 0.41 & 17.7 & 11.0 & -56.3 & -1.5 & 0.2\\
       190.63810 &       -2.4297249 &        315.32404 &        62.568188 & 19.04 & 1.02 & -0.21 & 0.25 & 67.7 & 6.0 & 4.4 & -1.0 & 0.5\\
       192.20784 &       -2.4226168 &        358.57367 &        48.349971 & 18.96 & 1.22 & -0.14 & 0.80 & 9.0 & 18.0 & 17.3 & -10.0 & -10.0\\
       172.36979 &       -1.4338628 &        272.87946 &        68.009004 & 18.48 & 1.24 & -0.15 & 0.89 & 22.3 & 8.0 & -55.1 & -1.4 & 0.1\\
       126.14404 &        46.955835 &        339.48188 &        55.003659 & 18.87 & 1.10 & -0.12 & 0.06 & 13.8 & 14.0 & -20.2 & -10.0 & -10.0\\
       123.44799 &        46.625513 &        334.21365 &        57.176227 & 18.67 & 1.13 & -0.03 & 0.19 & 17.5 & 14.0 & -24.7 & -1.6 & 0.0\\
       140.29301 &        58.048908 &        351.40623 &        50.936195 & 19.12 & 1.33 & -0.10 & 0.05 & 62.5 & 7.0 & 53.1 & -1.4 & 0.1\\

\hline
\end{tabular} 
\centering
\label{cleansample_table}
\end{scriptsize}
\end{table*}

\section{Discussion and Conclusions}

A number of previous works have explored the properties of the Sgr tidal
debris using different stellar tracers, such as main-sequence turnoff stars, M
giants, subgiant stars, or BHB stars.  In this paper, we have presented a
discussion of the structure and properties of the Sagittarius stellar
stream using candidate BHB stars selected from the SDSS coverage of the
North Galactic Cap.  BHB stars are in many senses an excellent tracer of tidal structure: they are luminous and can be traced to $>100$kpc
distances from the Sun with current surveys; they are good standard candles
with $\sim 5$\% accurate distances; and although they are rather sparse compared to other stellar populations, they are still quite numerous in the Sgr tidal stream (with exception of the closest part of the leading arm as shown in Figure \ref{msto}).
Recently, there have been a number of Sgr stellar stream studies \citep[e.g.,][]{Yanny2009,NiedersteOstholt2010,Correnti2010} also based on various stellar populations, some including BHB stars in SDSS.

In contrast to those studies, we entirely focus on BHB stars for the analysis, using other data only to relate our distance scale to other distance indicators.
We attempt to use as pure a sample of BHB star candidates as is
possible for our analysis, minimizing to the greatest extent possible the high levels of contamination seen in earlier studies. For charting out the global structure of the
Sgr tidal stream we make use of a method that selects BHB stars from SDSS
imaging data using a spectroscopic training set to isolate areas of $ugr$
color space that give a sample that should consist of $\sim 70$\% BHB stars. This method does not just make binary acceptance or rejection decisions based on the position in color-space, but assign probabilities to the stars based on their position in color space. In our analysis we mostly reject stars with probabilities $<50\%$ and make use of the probability information for the remaining sample by weighting the individual stars by their probabilities.

We evaluate the precision of our distance determination through comparison with distance measurements of known clusters and dwarf spheroidals. We see an offset in the mean values of $\sim 4$\% for the literature values, and a distance variance of $\sim 5$\%. Comparison to the M giants in the Sgr orbital plane implies that the M giant distances should be revised upward by 8\% or 12\% when compared with our BHB scale or the cluster distance scale in the literature, respectively. The offset to the previously adopted M giant distance scale is expected to propagate through to the models built to match the M giant observations \citep[e.g.,][]{Sag2MASS4,LM2010a}.
When studying the kinematics of the Sgr tidal tail, we focus
on a sample of stars with SDSS DR7 spectroscopy classified as BHB stars
using the method of \citet{Sirko2004} and \citet{Xue2008}; this sample should be
$>90$\% BHB stars.

With these samples, we focus on four Sgr stream issues
that are not well-explored in the literature: a possible extension of the
trailing Sgr debris stream, the line-of-sight thickness of the leading
tail, the bifurcation of the leading arm and the heliocentric distances of the two branches in BHB stars and the isolation of a small sample of high-probability Sgr member
stars.

Using the photometric sample with a $>50\%$ BHB probability limitation, we identify a possible
extension to the trailing tail of the Sgr debris stream to $60-80$ kpc.  The densest part
of this feature, which coincides spatially with the globular cluster NGC
2419, was previously argued to be associated with the Sgr trailing arm by
\citet{Newberg2003}. Our BHB star maps confirm a weak overdensity which may
be the extension of this arm back towards the Milky Way, which was also seen by \citet{Newberg2003,Newberg2007}. We estimate the significance of this feature to be around 3.8 $\sigma$ as compared to random selections of the same area within a region spanning the angular range and distance of the proposed trailing arm. A concentration in this region, which is claimed to be associated with the trailing arm of Sgr was also found by \citet{Sharma2010} in 2MASS M giants through a group finding technique. Such a feature is
expected qualitatively by models of Sgr disruption. Quantitatively, all models
predict that this `returning' segment of the trailing arm should be closer
to the Sun along these lines of sight. Yet, BHB stars are {\it not} observed at
these predicted distances, and this
tension would be resolved if one instead interpreted this distant
overdensity as this predicted part of the trailing Sgr arm and acknowledges a discrepancy between the positions predicted by the  models and the observed location. In this context it is worth noting that recently \citet{Correnti2010} reported an overdensity in Red Clump stars consistent with the predicted trailing arm location in the prolate models in a range of $220^{\circ} \lesssim \Lambda_{\odot} \lesssim 290^{\circ}$. Owing to confusion between Sgr and Virgo overdensity debris at the distance ranges probed by \citet{Correnti2010}, we were unable to confirm or refute this feature. If their feature is indeed correctly interpreted as trailing arm debris, we would suggest the feature identified here may be another, more distant wrap of trailing arm debris from an earlier close passage of Sgr.

We use the $>50\%$ probability sample to characterize the leading arm of the
Sgr stellar stream more closely by measuring the line-of-sight thickness
and selecting a high-probability sample of member stars.  We find a mean
thickness of $\sim 3$\ kpc, after accounting for distance uncertainties in
the BHB stars\footnote{Note that we only take into account the uncertainties estimated on single metallicity populations and not the additional uncertainty introduced by having a variety of metallicities. This would cause an underestimation of the distance uncertainty, which would result in an overestimation of the thickness of the arm.}, comparable to the projected width of Sgr on the sky. These measurements are in a similar range as those given by \citet{Correnti2010} in Red Clump stars (their Table 2), when the assumed overestimation by a factor of $\gtrsim2$ which is introduced by their measurement method is taken into account. Inspired by the clear appearance of the leading
arm in position (and velocity), we use this measurement of the line-of-sight thickness to select a sample of highly likely stream stars from the
spectroscopic SDSS sample. We choose stars within 2$\sigma$ of the stream
in line-of-sight distance. This subsample of stars shows a clear
overdensity in velocity space, which matches model predictions reasonably well. This strengthens the results of \citet{Yanny2009} who showed that BHB star candidates in the area selected to represent the spatial position of the leading arm in K/M-giants were overdense in velocity space. They find a similar trend in velocity space, but with far more outliers, probably due to the higher level of contamination in the BHB star sample and the broader selection box for the leading arm.  From our spatially selected sample, we measure an average velocity dispersion
of Sgr stars of 37\,km\,s$^{-1}$.  We further select stars within 1$\sigma$
of this velocity overdensity to be in our 'clean' sample of $\sim 70$ Sgr
BHB stars; such a sample suffers from the inhomogeneous angle coverage and
distance limitations inherent to spectroscopically-selected SDSS BHB stars,
but has the advantage of high fidelity. 

Using the spatially selected Sgr BHB star sample we examine the observed bifurcation of the Sgr stream on the sky and its implications for the distances of the two branches. Different distances of the two branches of the stream have been reported by e.g., \citet{Belokurov2006}, \citet{Yanny2009} and \citet{NiedersteOstholt2010} and are also predicted by models \citep{Penarrubia2010b}. This bifurcation is of particular interest for the understanding of the formation history of the Sgr stream. It has been a challenge for models to reproduce and explain the origin of this feature. It was proposed to have originated from wraps of different age \citep{Fellhauer2006}, but also intrinsic properties of the progenitor were offered as explanations. Recently, \citet{Penarrubia2010b} presented a model in which this bifurcation is reproduced when a rotating disk galaxy is assumed as the progenitor. There is no indication of a spatial bifurcation on the sky in our BHB star sample therefore we use MSTO stars to define the two branches on the sky. We measure the mean and width of the two parts along the leading arm and find only a small 1-2 kpc offset between the two branches over a wide orbital angle range and no clear systematic separation in the sense that one is always clearly closer in than the other \citep[c.f. ][]{Yanny2009,Correnti2010}. \citet{Yanny2009} reported a trend by visual impression that the branch at higher galactic latitude tends to be at closer distances than the branch at lower galactic latitude. \citet{Correnti2010} found similar to slightly higher offsets between the distances of the two branches compared to this study, but here again no clear trend is seen for one branch being always at closer distances than the other. Although the actual values for mean distances of the two branches in our data set seem to be quite sensitive to the precise location of the separation cut on the sky, we clearly do not see a separation on the level shown in \citet{NiedersteOstholt2010} for the same range of positions along the stream or predicted by the models \citep{Penarrubia2010b}. It is worth noting that this distance separation appears here to be much smaller than the separation of the two branches on the sky which is at a $10^{\circ}$ level. This strong discrepancy between the small line-of-sight separation and the much larger separation perpendicular to that might be challenging to reproduce in the models.

\acknowledgements
We thank the referee for helpful suggestions. We thank Nicolas Martin for assistance with the 2MASS sample and Eva Grebel for useful discussions.
C.\ R.\ was supported by the Emmy Noether Programme of the Deutsche Forschungsgemeinschaft, and is a member of the Heidelberg International Max Planck Research School program. E.\ F.\ B.\ acknowledges NSF grant AST 1008342. X.-X.\ X.\ acknowledges the support of the Max-Planck-Institute for
Astronomy, and the National Natural Science Foundation of China
(NSFC) under Nos. 10821061, 10876040, and 10973021.

Funding for the SDSS and SDSS-II has been provided by the Alfred P. Sloan Foundation, the Participating Institutions, the National Science Foundation, the U.S. Department of Energy, the National Aeronautics and Space Administration, the Japanese Monbukagakusho, the Max Planck Society, and the Higher Education Funding Council for England. The SDSS Web site is http://www.sdss.org/.

The SDSS is managed by the Astrophysical Research Consortium for the Participating Institutions. The Participating Institutions are the American Museum of Natural History, Astrophysical Institute Potsdam, University of Basel, University of Cambridge, Case Western Reserve University, University of Chicago, Drexel University, Fermilab, the Institute for Advanced Study, the Japan Participation Group, Johns Hopkins University, the Joint Institute for Nuclear Astrophysics, the Kavli Institute for Particle Astrophysics and Cosmology, the Korean Scientist Group, the Chinese Academy of Sciences (LAMOST), Los Alamos National Laboratory, the Max-Planck-Institute for Astronomy (MPIA), the Max-Planck-Institute for Astrophysics (MPA), New Mexico State University, Ohio State University, University of Pittsburgh, University of Portsmouth, Princeton University, the United States Naval Observatory, and the University of Washington.

\end{document}